\documentclass[12pt]{article}

\usepackage{cite}
\usepackage{amsmath}
\usepackage{amssymb}
\usepackage{bm}
\usepackage{graphicx}
\numberwithin{equation}{section}

\DeclareMathOperator{\diag}{diag}
\DeclareMathOperator{\Det}{Det}
\DeclareMathOperator{\Ai}{Ai}

\newcounter{aff}

\begin{document}
\begin{titlepage}
\begin{flushright}
{\footnotesize NITEP 127, OCU-PHYS 554}
\end{flushright}
\begin{center}
{\Large\bf
Duality Cascades and Affine Weyl Groups
}\\
\bigskip\bigskip
{
Tomohiro Furukawa\footnote{\tt furukawa@tx.osaka-cu.ac.jp},
Kazunobu Matsumura\footnote{\tt kmatsumura@jr.osaka-cu.ac.jp},
Sanefumi Moriyama\footnote{\tt sanefumi@osaka-cu.ac.jp},
Tomoki Nakanishi\footnote{\tt nakanishi@ka.osaka-cu.ac.jp}
}\\
\bigskip
${}^*{}^\dagger{}^\ddagger{}^\S$\,{\it Department of Physics, Graduate School of Science,}\\
{\it Osaka City University, Sumiyoshi-ku, Osaka 558-8585, Japan}\\[3pt]
${}^{\ddagger}$\,{\it Nambu Yoichiro Institute of Theoretical and Experimental Physics (NITEP),}\\
{\it Osaka City University, Sumiyoshi-ku, Osaka 558-8585, Japan}\\[3pt]
${}^\ddagger$\,{\it Osaka City University Advanced Mathematical Institute (OCAMI),}\\
{\it Osaka City University, Sumiyoshi-ku, Osaka 558-8585, Japan}
\end{center}

\begin{abstract}
Brane configurations in a circle allow subsequent applications of the Hanany-Witten transitions, which are known as duality cascades.
By studying the process of duality cascades corresponding to quantum curves with symmetries of Weyl groups, we find a hidden structure of affine Weyl groups.
Namely, the fundamental domain of duality cascades consisting of all the final destinations is characterized by the affine Weyl chamber and the duality cascades are realized as translations of the affine Weyl group, where the overall rank in the brane configuration associates to the grading operator of the affine algebra.
The structure of the affine Weyl group guarantees the finiteness of the processes and the uniqueness of the endpoint of the duality cascades.
In addition to the original duality cascades, we can generalize to the cases with Fayet-Iliopoulos parameters.
There we can utilize the Weyl group to analyze the fundamental domain similarly and find that the fundamental domain continues to be the affine Weyl chamber.
We further interpret the Weyl group we impose as a ``half'' of the Hanany-Witten transition.
\end{abstract}

\end{titlepage}

\tableofcontents

\section{Introduction}

The Hanany-Witten (HW) brane transitions \cite{HW} have played a crucial role in understanding supersymmetric gauge theories.
It is known that when an NS5-brane and a D5-brane placed perpendicularly at different positions in a line move across each other, a D3-brane is generated in the interval between the two 5-branes.
Later, a generalization of this brane configuration revives in the construction of the worldvolume theory of M2-branes, the Aharony-Bergman-Jafferis-Maldacena (ABJM) theory \cite{ABJM}.
On one hand, the HW transitions work generically for all brane configurations.
On the other hand, generalizations of the ABJM theory relate to the idea of quantum curves and enjoy large symmetries of Weyl groups of exceptional algebras.
We may gain new insights on both brane transitions and quantum curves by relating them directly.
Indeed, in studying the partition functions of the generalized ABJM theories, recently new types of brane transitions were proposed \cite{KM,FMN,Kubo}.
In this paper we shall turn to subsequent applications of the HW transitions in a circle known as duality cascades and study the processes with Weyl groups.

The ABJM theory \cite{ABJM,HLLLP2,ABJ} is the ${\cal N}=6$ supersymmetric Chern-Simons theory with gauge group $\text{U}(N_1)_k\times\text{U}(N_2)_{-k}$ (where the subscripts denote the Chern-Simons levels with $k >0$) and two pairs of bifundamental matters.
This theory describes the worldvolume of $\min(N_1,N_2)$ M2-branes and $|N_2-N_1|$ fractional M2-branes on a background ${\mathbb C}^4/{\mathbb Z}_k$.
One of the most important non-trivial checks of the description is that the free energy of the ABJM theory reproduces the degrees of freedom $N^{\frac{3}{2}}$ for $N$ M2-branes, as predicted from the gravity side \cite{KT}.
Namely, using the localization techniques \cite{P,KWY}, it is known that the partition function on $S^3$ defined with the infinite-dimensional path integral reduces to a finite-dimensional matrix integration (ABJM matrix model) and reproduces the $N^{\frac{3}{2}}$ behavior from the 't Hooft expansion \cite{DMP1} (with the Chern-Simons levels $k$ fixed by suitable reinterpretations).
It was further found that, for $N$ M2-branes, all the perturbative corrections are summed up to the Airy function \cite{FHM}, reducing to $\log\Ai(N)\sim N^{\frac{3}{2}}$ in the large $N$ limit.
The 't Hooft expansion further enables to uncover two types of non-perturbative effects, the worldsheet instantons \cite{DMP1} and the membrane instantons \cite{DMP2}.

It is interesting to observe that the integral representation of the Airy function closely resembles the inverse transformation from the grand potential $J_k(\mu)$ to the partition function;
\begin{align}
\Ai(N)=\int\frac{d\mu}{2\pi i}e^{\frac{1}{3}\mu^3-\mu N},\quad
Z_k(N)=\int\frac{d\mu}{2\pi i}e^{J_k(\mu)-\mu N}.
\end{align}
For this reason it seems more natural to study M2-branes in the grand canonical ensemble instead of the original canonical ensemble.
These observations open up new venues for the matrix model in an epoch-making paper \cite{MP}, which proposes to move to the grand canonical ensemble
\begin{align}
\Xi_k(z)=\sum_{N=0}^\infty z^NZ_k(N).
\end{align}
Then, the grand canonical partition function is given by the Fredholm determinant
\begin{align}
\Xi_k(z)=\Det(1+z\widehat H^{-1}),
\end{align}
where the quantum-mechanical spectral operator is given by $\widehat H=\widehat{\cal Q}\widehat{\cal P}$, with $(\widehat{\cal Q},\widehat{\cal P})$ being the operators defined by
\begin{align}
\widehat{\cal Q}=\widehat Q^{\frac{1}{2}}+\widehat Q^{-\frac{1}{2}},\quad
\widehat{\cal P}=\widehat P^{\frac{1}{2}}+\widehat P^{-\frac{1}{2}},\quad
\widehat Q\widehat P=q\widehat P\widehat Q.
\label{QP}
\end{align}
The parameter $q$ is identified to the level $k$ through $q=e^{i\hbar}$ and $\hbar=2\pi k$.
These operators may look more familiar if we express them as the exponentiated canonical operators
$(\widehat Q,\widehat P)=(e^{\widehat q},e^{\widehat p})$ satisfying the canonical commutation relation $[\widehat q,\widehat p]=i\hbar$.
Note that the spectral operator $\widehat H$ takes the form of the defining equation for ${\mathbb P}^1\times{\mathbb P}^1$ after suitable redefinitions, which leads to the idea of quantum curves \cite{MiMo,ACDKV}.
For a general setup with more 5-branes, it is known that we can directly translate $(1,k)$5-branes and NS5-branes in brane configurations into the operators $\widehat{\cal Q}$ and $\widehat{\cal P}$ in the spectral operator $\widehat H$ respectively in the reverse order \cite{MP,MN1,MN2}.
These correspond to various ${\cal N}=4$ supersymmetric Chern-Simons theories of circular quivers with more nodes of equal ranks, where levels are determined by the types of adjacent 5-branes (see for example \eqref{ks} below).

With this formalism, we can proceed to study the WKB expansion in the large $\mu$ limit (corresponding to the large $N$ limit).
Since the large $\mu$ limit of the grand potential $J_k(\mu)$ defined in
\begin{align}
\Xi_k(e^\mu)=\sum_{n=-\infty}^\infty e^{J_k(\mu+2\pi in)},
\end{align}
is found from the phase space area by taking the derivative, it always behaves as $J_k(\mu)=C\mu^3/3$ with a coefficient $C$ after integrations, which guarantees the partition function to be the Airy function.
In this sense the spectral operator encodes the keys to uncover the M2-branes if we regard the $N^{\frac{3}{2}}$ behavior or the Airy function as characteristics of multiple M2-branes.

The Fermi gas formalism further leads us to compute some exact values of the ABJM matrix model \cite{HMO1,PY} which enables to uncover the whole structure of the non-perturbative effects \cite{HMO2,CM,HMO3}.
It was finally found \cite{HMMO} that the non-perturbative effects of the grand potential are given by the free energy of topological strings and the derivative of its refinements on local ${\mathbb P}^1\times{\mathbb P}^1$.
Interestingly, by removing the role of the ABJM matrix model, it was proposed \cite{GHM1} that generally the Fredholm determinant with spectral operators for curves is given by the free energy of topological strings on the correspondingly local geometries.

All these progresses suggest that quantum curves suitably describe the multiple M2-branes.
For this reason it is desirable to understand the quantum curves better.
Besides, it is known that the classical cousins of the spectral operators of genus one reduce to the defining equations for the del Pezzo geometries, which enjoy large symmetries of Weyl groups of exceptional algebras.
It is then interesting to figure out how the geometries and the symmetries of exceptional Weyl groups are lifted quantum-mechanically to deepen our understanding of M2-branes.
It turns out that the group-theoretical structure of the quantum curves works crucially for understanding better the generalized ABJM matrix models \cite{HM,MN1,MN2,MN3,HHO,GKMR,CGM,MNN} in \cite{MNY,KMN,FMS,M}, where mirror maps and BPS indices \cite{HKP} for the matrix models are characterized by representations of the corresponding groups.

The quantum curves enjoying the Weyl group symmetries are known to appear in the context of $q$-Painlev\'e equations as conserved curves for autonomous systems \cite{Sakai,KNY}.
For general non-autonomous ones, it is known that the time evolutions in $q$-Painlev\'e equations are generated by the translations appearing in the affine Weyl groups.
For this reason, it is interesting to work out the affine Weyl groups from the quantum curves explicitly.
The $E_8$ quantum curve of the highest rank was constructed in \cite{M} and the complete quantum affine Weyl group including tau variables was worked out recently in \cite{MY}.

We also gain insights for brane transitions from the quantum curves.
Especially for partition functions, we can relate directly the space of rank deformations in brane configurations to the parameter space of quantum curves and translate symmetries of the Weyl group into brane transitions \cite{KM,FMN}.
Namely, for brane configurations, even starting from the partition functions without rank deformations, we can apply the HW transitions to exchange 5-branes into the standard order and read off the rank deformations.
For quantum curves constructed by \eqref{QP}, we easily read off the parameters of the curves.
Then, a direct relation between brane configurations and quantum curves is found by comparison.
In this sense, the symmetries of the exceptional Weyl groups are translated to the symmetries of brane configurations and it is possible to find out new brane transitions unknown previously from large symmetries of the exceptional Weyl groups.
Note that, as pointed out in \cite{KM}, in comparison we need to fix an interval where the HW transitions are forbidden.
Since fixing an interval is similar to the idea of fixing a reference frame in classical mechanics or a local patch in manifolds, we refer to the interval as the reference.

The idea of fixing the reference interval originates from the study of the grand canonical partition functions.
In fact, in summing up the partition functions in the grand canonical ensemble, it was pointed out \cite{MNN} that the power of the fugacity has to be fixed to the rank in one specific interval and the starting point of the summation needs to be chosen to be the lowest possible rank.
Also, in describing M2-branes using the spectral operators $\widehat H$ in the so-called closed string formalism \cite{H1,KMZ,MS2,MN5,closed}, it is desirable to separate fractional M2-branes from full-fledged M2-branes and regard the fractional M2-branes as small fluctuations to be integrated out.
From these viewpoints, it is natural to expect the reference rank to be the lowest among all the ranks.

Furthermore, it was pointed out that quantum curves can be used in the study of duality cascades.
The duality cascade \cite{KS} is a series of subsequent duality transformations in field theories, which is realized by the HW transitions in brane configurations in a circle.
For the ABJM case with one NS5-brane and one $(1,k)$5-brane, the HW transition implies
\begin{align}
\langle N\bullet N+M\circ\rangle=\langle N\circ N-M+k\bullet\rangle,
\end{align}
where we express the NS5-brane and the $(1,k)$5-brane as $\bullet$ and $\circ$ respectively, fill in the numbers of D3-branes $(N_1,N_2)=(N,N+M)$ in between and indicate the reference interval by the endpoint of the bracket.
We assume that both the overall rank $N$ and the relative rank $M$ to be non-negative.
If $M\le k$, we can repeat the transition and return to the original brane configuration.
However, if $M>k$, apparently the lowest rank changes, $N>N-M+k$, and it is more suitable to change the reference rank $N$ into $N-M+k$ by cyclic rotations and consider the brane configuration $\langle N-M+k\bullet N\circ\rangle$.
Namely, in the field-theoretical language, the original gauge group $\text{U}(N)_k\times\text{U}(N+M)_{-k}$ is transformed into $\text{U}(N-M+k)_k\times\text{U}(N)_{-k}$.
Apparently, the configuration returns to the original one with both the parameters $(N,M)$ decreasing as in
\begin{align}
(N,M)\to(N-M+k,M-k).
\label{NMchange}
\end{align}
Returning to the brane configuration with 5-branes of the same order, we can continue to apply the HW transitions and repeat the same process to decrease the parameters $(N,M)$.
Since $M$ keeps on decreasing, finally $M$ satisfies $0\le M\le k$.
It was proposed that, when one of the ranks is negative, the brane configuration is non-supersymmetric \cite{HW,AHHO}.
Whether the brane configuration is supersymmetric then depends on which of $N$ and $M$ decreases faster.
Here we restrict ourselves to the cases that $N$ is large enough so that the configuration is always supersymmetric.
Recently, it was shown \cite{HK} that the process of duality cascades is consistent with the computation of the quantum curves for the partition functions.

Though it was not stated explicitly before as far as we know, we can summarize the working hypothesis of duality cascades for a system with multiple 5-branes as follows.
\begin{enumerate}
\item Consider a brane configuration in a circle with perpendicular 5-branes tilted relatively and D3-branes in each interval.
\item Specify one of the lowest ranks (the lowest numbers of D3-branes) as the reference.
\item Apply the HW transitions to exchange the 5-branes arbitrarily without crossing the reference.
\item When ranks lower than the reference appear, relabel one of the lowest ranks as the reference.
\item Repeat the last two steps.
\end{enumerate}

In retrospect of all these studies, it is natural to ask the following questions.
\begin{itemize}
\item
Although the quantum curves enjoy symmetries of Weyl groups, it is unclear how the affine Weyl groups of $q$-Painlev\'e equations work in the brane configurations.
Also, it is interesting to elaborate the time evolutions of $q$-Painlev\'e equations in brane configurations (see \cite{GHM2,BGT}).
\item
It is also unclear to us whether the working hypothesis of duality cascades is physically well-defined.
Namely, we would like to clarify whether the processes of duality cascades will always end and whether the endpoint is unique if two different processes of the HW transitions are applied.
This is particularly unclear if we increase the numbers of 5-branes with more relative ranks, since, unlike the ABJM case \eqref{NMchange} with only one relative rank, multiple relative ranks do not necessarily change monotonically.
\item
Assuming the duality cascades always end and the endpoint is unique, we may collect all of the endpoints in the duality cascades.
We refer to the collection of endpoints as {\it the fundamental domain of duality cascades} since each endpoint serves as the representative of all the supersymmetric brane configurations appearing in the duality cascades.
It is also interesting to clarify the geometrical properties of the fundamental domain such as connectedness, convexity and compactness.
\end{itemize}

For example, consider the brane configuration $\langle 12\stackrel{2}{\bullet}25\stackrel{1}{\bullet}34\stackrel{3}{\circ}25\stackrel{4}{\circ}\rangle$ with the level $k=3$, where we distinguish 5-branes by labels.
Following the working hypothesis of duality cascades, we can move to $\langle 12\stackrel{1}{\bullet}21\stackrel{3}{\circ}15\stackrel{4}{\circ}5\stackrel{2}{\bullet}\rangle$ by the HW transitions and change references to $\langle 5\stackrel{2}{\bullet}12\stackrel{1}{\bullet}21\stackrel{3}{\circ}15\stackrel{4}{\circ}\rangle$ by rotations since lower ranks appear.
Then we can move again to $\langle 5\stackrel{4}{\circ}1\stackrel{2}{\bullet}5\stackrel{1}{\bullet}11\stackrel{3}{\circ}\rangle$ and arrive at $\langle 1\stackrel{2}{\bullet}5\stackrel{1}{\bullet}11\stackrel{3}{\circ}5\stackrel{4}{\circ}\rangle$ by changing references.
Alternatively, from the same initial brane configuration, we can move to $\langle 12\stackrel{4}{\circ}5\stackrel{2}{\bullet}15\stackrel{1}{\bullet}21\stackrel{3}{\circ}\rangle$, change references to $\langle 5\stackrel{2}{\bullet}15\stackrel{1}{\bullet}21\stackrel{3}{\circ}12\stackrel{4}{\circ}\rangle$ and move again to $\langle 5\stackrel{1}{\bullet}11\stackrel{3}{\circ}5\stackrel{4}{\circ}1\stackrel{2}{\bullet}\rangle$, which surprisingly reduces to the same final brane configuration by rotations (see table \ref{cascadeexample}).
Namely, apparently it is unclear how the two processes of duality cascades reduce to the same final brane configuration.
It is also unclear at the first sight whether we can continue the duality cascades for $\langle 1\stackrel{2}{\bullet}5\stackrel{1}{\bullet}11\stackrel{3}{\circ}5\stackrel{4}{\circ}\rangle$ or not. 
Note that, of course, all of these brane configurations can be translated directly into the field-theoretical language.

\begin{table}
\begin{center}
\begin{tabular}{c}
$\rotatebox[origin=c]{30}{=}\;
\langle 12\stackrel{1}{\bullet}21\stackrel{3}{\circ}15\stackrel{4}{\circ}5\stackrel{2}{\bullet}\rangle
\to\langle 5\stackrel{2}{\bullet}12\stackrel{1}{\bullet}21\stackrel{3}{\circ}15\stackrel{4}{\circ}\rangle
=\langle 5\stackrel{4}{\circ}1\stackrel{2}{\bullet}5\stackrel{1}{\bullet}11\stackrel{3}{\circ}\rangle
\;\rotatebox[origin=c]{-30}{$\to$}$\\[2pt]
$\langle 12\stackrel{2}{\bullet}25\stackrel{1}{\bullet}34\stackrel{3}{\circ}25\stackrel{4}{\circ}\rangle\hspace{9cm}
\langle 1\stackrel{2}{\bullet}5\stackrel{1}{\bullet}11\stackrel{3}{\circ}5\stackrel{4}{\circ}\rangle$\\[2pt]
$\rotatebox[origin=c]{-30}{=}\;
\langle 12\stackrel{4}{\circ}5\stackrel{2}{\bullet}15\stackrel{1}{\bullet}21\stackrel{3}{\circ}\rangle
\to\langle 5\stackrel{2}{\bullet}15\stackrel{1}{\bullet}21\stackrel{3}{\circ}12\stackrel{4}{\circ}\rangle
=\langle 5\stackrel{1}{\bullet}11\stackrel{3}{\circ}5\stackrel{4}{\circ}1\stackrel{2}{\bullet}\rangle
\;\rotatebox[origin=c]{30}{$\to$}$
\end{tabular}
\end{center}
\caption{Two different processes of duality cascades.
Both of them reduce to the same final brane configuration.}
\label{cascadeexample}
\end{table}

In this paper, we shall answer the above questions by pointing out that the working hypothesis of duality cascades matches well with the affine Weyl group associated to quantum curves.
We mainly choose the brane configurations corresponding to the $D_5$ or $E_7$ curves;
unfortunately those corresponding to the $A_1$ curve (the ABJM case) \eqref{NMchange} are too simple to convince us of the structure.
One of the main claims of this paper is that the fundamental domain of duality cascades in the Weyl chamber is nothing but {\it the affine Weyl chamber}.
By this identification, duality cascades are equivalent to the geometrical process of transforming the relative ranks by translations and decreasing the overall rank, which is nothing but an element of the affine Weyl group.
This guarantees the physical properties expected for the duality cascades, the finiteness of the processes and the uniqueness of the endpoint, given $N$ large enough.

We first start our analysis in the space without deformations of Fayet-Iliopoulos (FI) parameters.
We study the fundamental domain of duality cascades by requiring the reference rank to continue to be the lowest in all of the HW transitions and understand that the duality cascades are realized by translations in the parameter space of quantum curves.
It is known that for these cases without FI parameters the asymptotic values in the opposite regions always coincide with each other.
Due to this reason, the $D_5$ and $E_7$ Weyl groups are subject to the ${\mathbb Z}_2$ folding and reduce respectively to the $B_3$ and $F_4$ Weyl groups.
Then, we introduce the deformation of FI parameters for quantum curves with full parameters and study the same problem of the fundamental domain by requiring the inequalities obtained from the Weyl group.
To understand the imposed symmetries of the Weyl group, we further rewrite the transformations of the Weyl group in terms of brane configurations.

Interestingly, relations between duality cascades and affine Weyl groups were discussed previously in \cite{CFIKV,F,HRW} in other realizations of field theories by string theories where D-brane probes wrap cycles of local Calabi-Yau manifolds.
Comparatively, our setup of brane configurations consists only of flat branes in the flat spacetime and enjoys symmetries of the $B_3$ and $F_4$ Weyl groups without deformations.
We expect that the simple structure may lead to many generalizations of this idea even without symmetries of Weyl groups.

This paper is organized as follows.
We study the cases without deformations of FI parameters in the next section, while turn to the deformation in section \ref{FI}.
Finally we conclude with summaries and discussions on future directions.
Appendix is devoted to summaries of quantum curves and their symmetries.

\section{Fundamental domains and duality cascades}\label{FDDC}

In this section we shall study duality cascades corresponding to the quantum curves without deformations of FI parameters.
We start with the fundamental domain consisting of all the final destinations, and proceed to the processes of duality cascades themselves.

\subsection{Brane setup}

\begin{table}
\begin{center}
\begin{tabular}{c|ccccccc}
&0&1&2&6&3\quad 7&4\quad 8&5\quad 9\\\hline
D3&$-$&$-$&$-$&$-$&&&\\
NS5&$-$&$-$&$-$&&$-\quad\;\;$&$-\quad\;\;$&$-\quad\;\;$\\
$(1,k)$5&$-$&$-$&$-$&&$[3,7]_\theta$&$[4,8]_\theta$&$[5,9]_\theta$
\end{tabular}
\end{center}
\caption{Directions where branes are extending.
$(1,k)$5-branes extend in the directions of $3,4,5$ tilted respectively to the directions of $7,8,9$ by an angle, $\theta=\arctan k$, to preserve supersymmetries.}
\label{direction}
\end{table}

\begin{figure}[!t]
\centering\includegraphics[scale=0.6,angle=-90]{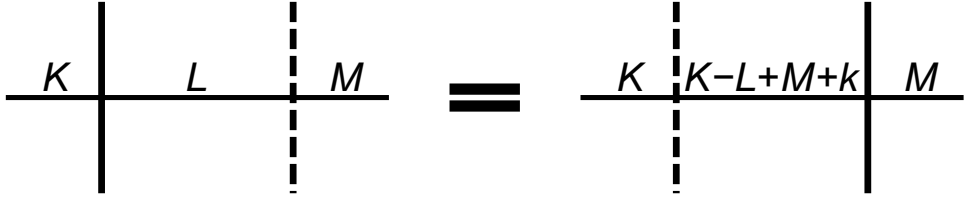}
\caption{Brane cartoon of the HW transition in \eqref{hw1}.
The horizontal direction denotes the direction 6, while the vertical direction denotes the directions of 3,4,5,7,8,9.
The horizontal line indicates the D3-branes and the two different vertical lines (solid and dashed) are the two different 5-branes (the NS5-brane and the $(1,k)$5-brane respectively).}
\label{hwmove}
\end{figure}

We first explain the setup of brane configurations.
The brane configurations we consider here are those with D3-branes compactified in a circle, where NS5-branes and $(1,k)$5-branes are placed perpendicularly and tilted relatively by an angle depending on $k$ to preserve supersymmetries.
The directions where branes are extending are depicted in table \ref{direction}.
These are generalizations from the original setup of \cite{HW} and play a crucial role in understanding the proposal on M2-branes in \cite{ABJM}.
Though in \cite{ABJM} the T-duality is taken and the brane configuration is lifted to M-theory, we stick to the IIB picture in this paper.
If we express NS5-branes and $(1,k)$5-branes by $\bullet$ and $\circ$ respectively and fill in the numbers of D3-branes in the interval, the standard HW transitions can be expressed by
\begin{align}
\cdots K\left[\begin{matrix}\bullet\\\circ\end{matrix}\right]L\left[\begin{matrix}\circ\\\bullet\end{matrix}\right]M\cdots
=\cdots K\left[\begin{matrix}\circ\\\bullet\end{matrix}\right]K-L+M+k\left[\begin{matrix}\bullet\\\circ\end{matrix}\right]M\cdots,
\label{hw1}
\end{align}
while the trivial ones of exchanging 5-branes of the same type are
\begin{align}
\cdots K\left[\begin{matrix}\bullet\\\circ\end{matrix}\right]L\left[\begin{matrix}\bullet\\\circ\end{matrix}\right]M\cdots=\cdots K\left[\begin{matrix}\bullet\\\circ\end{matrix}\right]K-L+M\left[\begin{matrix}\bullet\\\circ\end{matrix}\right]M\cdots.
\end{align}
Here double symbols of $\bullet$ and $\circ$ are in the same order.
See figure \ref{hwmove} for the brane cartoon of the HW transition.
It was noted in \cite{HW} that the transitions can be stated as conservation of 5-brane charges.
For example, for an NS5-brane, the HW transition is derived by requiring conservation of the charge
\begin{align}
Q_\text{RR}=-\frac{k}{2}\big((\#(1,k)5)_\text{L}-(\#(1,k)5)_\text{R}\bigr)+\big((\#\text{D3})|_\text{L}-(\#\text{D3})|_\text{R}\big),
\label{charge}
\end{align}
where $(\#(1,k)5)_\text{L/R}$ is the number of $(1,k)$5-branes located to the left/right of the NS5-brane, while $(\#\text{D3})|_\text{L/R}$ is the number of D3-branes ending on the NS5-brane from the left/right.
Apparently, we cannot distinguish left or right in a circle.
For this reason, in \cite{KM} it was proposed to specify the interval where the HW transitions are not applied.
Let us choose this interval as the reference and indicate it by the endpoint of the bracket $\langle\cdots\rangle$.
Correspondingly, the quantum curves are subject to similarity transformations rotating the exponentiated canonical operators $(\widehat{\cal Q},\widehat{\cal P})$ cyclically and can be ambiguous by uncritical uses of them.
This ambiguity is fixed by the reference simultaneously.
In the following discussions, we mainly have partition functions of the brane configurations in mind.
It is also interesting to understand to what extent the brane transitions hold with various physical excitations.

\subsection{$B_3$}\label{B3}

We first consider brane configurations corresponding to the $D_5$ quantum curve, which consist of two NS5-branes and two $(1,k)$5-branes placed standardly with 5-branes of the same types next to each other, ${\langle}{\bullet}{\bullet}{\circ}{\circ}{\rangle}$.
The corresponding gauge theories described by these brane configurations have gauge group $\text{U}(N_1)_k\times\text{U}(N_2)_0\times\text{U}(N_3)_{-k}\times\text{U}(N_4)_0$ and pairs of bifundamental matters connecting adjacent factors of the gauge group cyclically, where the ranks and the levels are subject to change due to the HW transitions.
As found in \cite{KM,FMN} and recapitulated in appendix \ref{appendixD5}, without further deformations, the brane configurations are parameterized by three relative ranks and enjoy symmetries of the $B_3$ Weyl group, which stems from the original $D_5$ Weyl group by applying the ${\mathbb Z}_2$ folding.

\subsubsection{Fundamental domain ($B_3$)}

\begin{table}[t!]
\begin{center}
\begin{tabular}{c||c}
$\langle\stackrel{2}{\bullet}\stackrel{1}{\bullet}\stackrel{3}{\circ}\stackrel{4}{\circ}\rangle$&inequality\\\hline
$\langle\stackrel{2}{\bullet}\cdot\stackrel{1}{\bullet}\stackrel{3}{\circ}\stackrel{4}{\circ}\rangle$&$M_1-M_2+M_3+k\ge 0$\\
$\langle\stackrel{1}{\bullet}\cdot\stackrel{2}{\bullet}\stackrel{3}{\circ}\stackrel{4}{\circ}\rangle$&$M_1+M_2-M_3+k\ge 0$\\
$\langle\stackrel{3}{\circ}\cdot\stackrel{2}{\bullet}\stackrel{1}{\bullet}\stackrel{4}{\circ}\rangle$&$-M_1-M_2-M_3+k\ge 0$\\
$\langle\stackrel{4}{\circ}\cdot\stackrel{2}{\bullet}\stackrel{1}{\bullet}\stackrel{3}{\circ}\rangle$&$-M_1+M_2+M_3+k\ge 0$\\
$\langle\stackrel{2}{\bullet}\stackrel{1}{\bullet}\cdot\stackrel{3}{\circ}\stackrel{4}{\circ}\rangle$&$2M_1+2k\ge 0$\\
$\langle\stackrel{2}{\bullet}\stackrel{3}{\circ}\cdot\stackrel{1}{\bullet}\stackrel{4}{\circ}\rangle$&$-2M_2+k\ge 0$\\
$\langle\stackrel{1}{\bullet}\stackrel{3}{\circ}\cdot\stackrel{2}{\bullet}\stackrel{4}{\circ}\rangle$&$-2M_3+k\ge 0$\\
$\langle\stackrel{2}{\bullet}\stackrel{4}{\circ}\cdot\stackrel{1}{\bullet}\stackrel{3}{\circ}\rangle$&$2M_3+k\ge 0$\\
$\langle\stackrel{1}{\bullet}\stackrel{4}{\circ}\cdot\stackrel{2}{\bullet}\stackrel{3}{\circ}\rangle$&$2M_2+k\ge 0$\\
$\langle\stackrel{3}{\circ}\stackrel{4}{\circ}\cdot\stackrel{2}{\bullet}\stackrel{1}{\bullet}\rangle$&$-2M_1+2k\ge 0$\\
$\langle\stackrel{2}{\bullet}\stackrel{1}{\bullet}\stackrel{3}{\circ}\cdot\stackrel{4}{\circ}\rangle$&$M_1-M_2-M_3+k\ge 0$\\
$\langle\stackrel{2}{\bullet}\stackrel{1}{\bullet}\stackrel{4}{\circ}\cdot\stackrel{3}{\circ}\rangle$&$M_1+M_2+M_3+k\ge 0$\\
$\langle\stackrel{2}{\bullet}\stackrel{3}{\circ}\stackrel{4}{\circ}\cdot\stackrel{1}{\bullet}\rangle$&$-M_1-M_2+M_3+k\ge 0$\\
$\langle\stackrel{1}{\bullet}\stackrel{3}{\circ}\stackrel{4}{\circ}\cdot\stackrel{2}{\bullet}\rangle$&$-M_1+M_2-M_3+k\ge 0$
\end{tabular}
\end{center}
\caption{Inequalities found by comparing the reference rank with various ranks (denoted by $\cdot$) in various brane configurations obtained from the HW transitions.}
\label{inequalityB3}
\end{table}

As in \cite{KM,FMN}, we label the relative ranks by
\begin{align}
&\langle N_1\stackrel{2}{\bullet}N_2\stackrel{1}{\bullet}N_3\stackrel{3}{\circ}N_4\stackrel{4}{\circ}\rangle
\nonumber\\
&=\langle N_0+M_2+M_3\stackrel{2}{\bullet}N_0+M_1+2M_3+k\stackrel{1}{\bullet}N_0+2M_1+M_2+M_3+2k\stackrel{3}{\circ}N_0+M_1+k\stackrel{4}{\circ}\rangle.
\label{M123}
\end{align}
Let us assume that, after applying the duality cascades subsequently, finally we arrive at the brane configurations without any possibilities of further cascades (which will be justified later in this subsection).
Then, the reference rank $N_0+M_2+M_3$ has to continue to be the lowest of all ranks, even when we apply the HW transitions preserving the reference arbitrarily.
By studying the condition of the lowest rank, we find 14 inequalities
\begin{align}
\pm M_1\le k,\quad\pm M_2\le k/2,\quad\pm M_3\le k/2,\quad\pm M_1\pm M_2\pm M_3\le k,
\label{ineqB3trivial}
\end{align}
where any combination of multiple signs may apply.
Since the ranks appearing in the HW transitions can be understood from charge conservation \eqref{charge} obtained from the types and the numbers of 5-branes located on the left or on the right, each of the 14 inequalities corresponds to the condition that the reference rank is lower than the rank with each combination of the 5-branes in between.
This gives $2^4-2=14$ inequalities by removing the trivial two: none and all, (see table \ref{inequalityB3}).
In the table, $\langle\stackrel{2}{\bullet}\stackrel{3}{\circ}\cdot\stackrel{1}{\bullet}\stackrel{4}{\circ}\rangle$, for example, is a short-hand notation where we abbreviate the numbers of D3-branes and indicate by $\cdot$ the rank which we compare with the reference rank.
In comparing ranks in $\langle\stackrel{2}{\bullet}\stackrel{3}{\circ}\cdot\stackrel{1}{\bullet}\stackrel{4}{\circ}\rangle$, we apply the HW transition for \eqref{M123} and change the order of 5-branes into
\begin{align}
\langle N_0+M_2+M_3\stackrel{2}{\bullet}N_0+M_1+2M_3+k\stackrel{3}{\circ}N_0-M_2+M_3+k\stackrel{1}{\bullet}N_0+M_1+k\stackrel{4}{\circ}\rangle.
\label{HWex}
\end{align}
Then, we obtain the inequality $N_0+M_2+M_3\le N_0-M_2+M_3+k$ from the comparison.

\begin{figure}[!t]
\centering\includegraphics[scale=0.7,angle=-90]{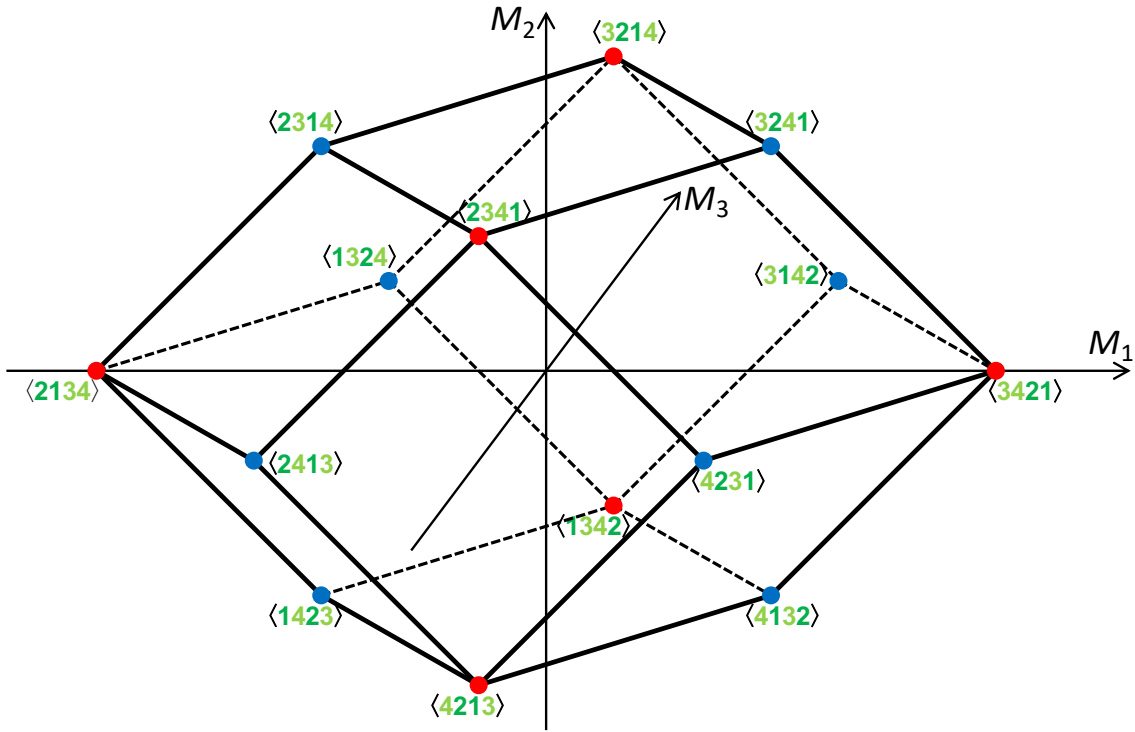}
\caption{The fundamental domain of duality cascades for brane configurations with two NS5-branes and two $(1,k)$5-branes.
The brane configurations without rank differences appear as vertices and we label the vertices by the brane configurations $\langle\cdots\rangle$ with $\bullet$ and $\circ$ omitted by denoting the labels in dark and light green.
Interestingly, the fundamental domain is the famous rhombic dodecahedron studied by Johannes Kepler.}
\label{domainB3}
\end{figure}

We usually refer to the region satisfying the inequalities \eqref{ineqB3trivial} as the fundamental domain of duality cascades, since we expect it to consist of all final destinations of duality cascades.
It was proposed \cite{HW,AHHO} that the brane configurations are non-supersymmetric when we encounter a negative rank in any interval.
Here we mainly consider the supersymmetric case where $N_0$ is large enough.
It is not difficult to find that the first two inequalities $\pm M_1\le k$ in \eqref{ineqB3trivial} are trivially guaranteed by summing up the four other inequalities $\pm(M_1\pm M_2\pm M_3)\le k$.
Then, the condition reduces to 12 inequalities
\begin{align}
\pm M_2\le k/2,\quad\pm M_3\le k/2,\quad\pm M_1\pm M_2\pm M_3\le k.
\label{ineqB3}
\end{align}
The fundamental domain is depicted pictorially in figure \ref{domainB3}.

It is interesting to observe that, after rescalings, the fundamental domain in figure \ref{domainB3} is the famous rhombic dodecahedron with 12 congruent rhombic faces and 14 vertices of two types: one with three faces intersecting (8 blue vertices) and one with four faces intersecting (6 red vertices).
The 12 faces correspond to the above 12 inequalities. 
The two types of vertices are understood as the brane configurations without rank differences (but with various orders of 5-branes).
This is natural from the following viewpoint.
In general vertices are defined as intersections of several hyperplanes originating from inequalities for each interval rank.
The brane configurations without rank differences are located marginally within the region for each inequality, which are nothing but the vertices.
The arguments, however, should be inspected carefully due to various symmetries of the brane configurations.
Here, the brane configurations with NS5-branes and $(1,k)$5-branes placed separately (${\langle}{\bullet}{\circ}{\bullet}{\circ}{\rangle}$ and ${\langle}{\circ}{\bullet}{\circ}{\bullet}{\rangle}$) correspond to the blue vertices with three faces intersecting, while those with 5-branes of the same types placed next to each other (${\langle}{\bullet}{\bullet}{\circ}{\circ}{\rangle}$, ${\langle}{\bullet}{\circ}{\circ}{\bullet}{\rangle}$, ${\langle}{\circ}{\bullet}{\bullet}{\circ}{\rangle}$ and ${\langle}{\circ}{\circ}{\bullet}{\bullet}{\rangle}$) correspond to the red ones with four faces intersecting.
For the blue vertices the three faces correspond respectively to the three inequalities for relative ranks, while for the red vertices the situation is more complicated due to the extra symmetries exchanging 5-branes of the same type.

\begin{figure}[!t]
\centering\includegraphics[scale=0.6,angle=-90]{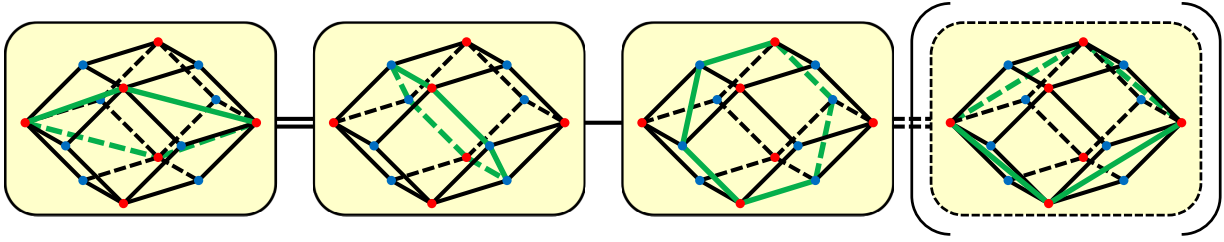}
\caption{Symmetries of the fundamental domain of duality cascades.
Symmetries are understood by the $B_3$ Weyl group generated from Weyl reflections $s_{12}=s_1s_2$, $s_3$, $s_4$ and $s_{50}=s_5s_0$, though the metric may not be orthonormal.
Each transformation is given explicitly in appendix \ref{appendixD5}.
The rightmost reflection $s_{50}$ is neither independent nor affine; it is just obtained naturally by folding the affine Dynkin diagram of $D_5$.}
\label{symmetryB3}
\end{figure}

Interestingly, it was found in \cite{KM} and further explained in \cite{FMN} that the brane configurations enjoy symmetries of the $B_3$ Weyl group.
As recapitulated in appendix \ref{appendixD5}, the appearance of the $B_3$ Weyl group can be understood from the ${\mathbb Z}_2$ folding of the affine Dynkin diagram of $D_5$.
Hence the fundamental domain also enjoys the same symmetries.
We can utilize the $B_3$ Weyl group to understand the fundamental domain better.

The Weyl group is generated by Weyl reflections $s_\alpha\lambda=\lambda-2\alpha(\alpha,\lambda)/(\alpha,\alpha)$ \eqref{reflection} on the parameters of relative ranks $\lambda$ associated to the simple roots $\alpha$.
For the current case, the parameters are given by $\lambda=(M_1,M_2,M_3)$ while the metric for the bilinear form and the simple roots ($\alpha_{12}$, $\alpha_3$ and $\alpha_4$) are given respectively by $g=\diag(1,2,2)$ and \eqref{B3roots}.
Pictorially, the reflections are given in figure \ref{symmetryB3}.
Note that the rightmost Weyl reflection $s_{50}=s_5s_0$ (combined from those in the $D_5$ Weyl group) in the figure is not independent but auxiliary.
Besides, it is not even the affine root.
We depict it in the figure only because that it appears naturally in the ${\mathbb Z}_2$ folding of the affine Dynkin diagram of $D_5$.
Due to this reason, the overall sign of the corresponding root $\alpha_{50}$ in \eqref{B3roots} is not fixed.
Of course, we can rescale the directions of $M_2$ and $M_3$ or introduce variables $M_\pm=M_2\pm M_3$ to normalize the metric $g=\diag(1,2,2)$.
This normalization explains the rescalings for the rhombic dodecahedron.

\begin{figure}[!t]
\centering\includegraphics[scale=0.5,angle=-90]{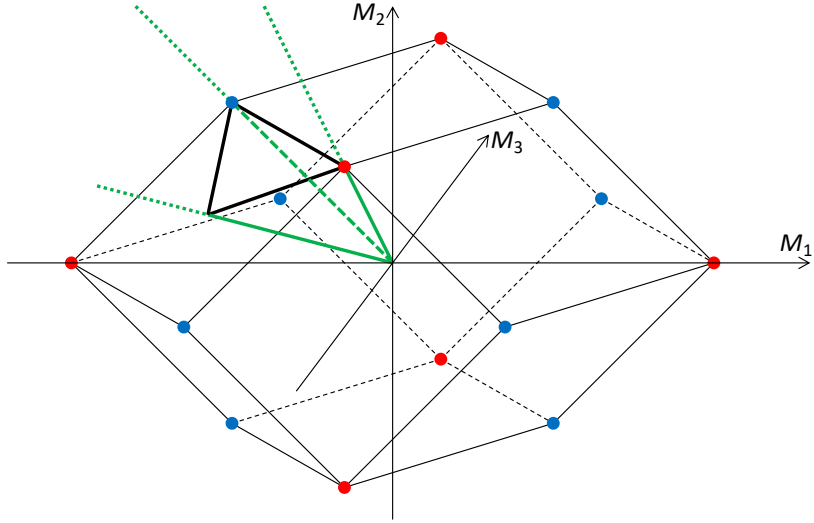}
\caption{The fundamental domain in the $B_3$ Weyl chamber \eqref{B3chamber} (surrounded by green lines).}
\label{chamberB3}
\end{figure}

Now let us study the fundamental domain using the $B_3$ Weyl group.
Since we already know that the brane configurations enjoy symmetries of the $B_3$ Weyl group, we can restrict our analysis to the $B_3$ Weyl chamber.
Namely, instead of investigating the whole fundamental domain, we can constrain ourselves to
\begin{align}
(\alpha_{12},\lambda)=M_2+M_3\ge 0,\quad
(\alpha_3,\lambda)=-M_1-M_2-M_3\ge 0,\quad
(\alpha_4,\lambda)=M_1+M_2-M_3\ge 0.
\label{B3chamber}
\end{align}
It is not difficult to study the intersections in the three-dimensional space from figure \ref{domainB3} and find the fundamental domain reduces to that in the $B_3$ Weyl chamber (depicted by the green cone in figure \ref{chamberB3}).
Reversely, starting with the fundamental domain in the $B_3$ Weyl chamber, we can obtain the original fundamental domain by applying the $B_3$ Weyl group.
Namely, the whole information of the fundamental domain reduces to the single inequality
\begin{align}
-M_1+M_2-M_3\le k,
\label{boundaryB3}
\end{align}
in \eqref{ineqB3}; all others are obtained from the $B_3$ Weyl group.

Although in three dimensions the intersection is geometrically intuitive, for later applications, we also comment on the analytical computation.
Among the three planes in \eqref{B3chamber}, two of them define an edge of the Weyl chamber.
We can then study intersection points between these edges and the boundaries of the fundamental domain, which are
\begin{align}
k\omega_{12}=(-k/2,k/2,0),\quad
k\omega_3/2=(-k/2,k/4,-k/4),\quad
k\omega_4=(0,k/2,-k/2),
\label{convexB3}
\end{align}
where $\omega_{12}$, $\omega_3$ and $\omega_4$ are the fundamental weights \eqref{B3fw} and the inverse comarks appear in the coefficients.
Since our fundamental domain is defined from inequalities, the fundamental domain is obviously convex containing these three points, all of which lie on the plane $-M_1+M_2-M_3=k$.
Since the inequality \eqref{boundaryB3} already appears in the condition \eqref{ineqB3}, the fundamental domain in the $B_3$ Weyl chamber is nothing but the convex hull of the three points \eqref{convexB3} and the origin.

\begin{figure}[!t]
\centering\includegraphics[scale=0.7,angle=-90]{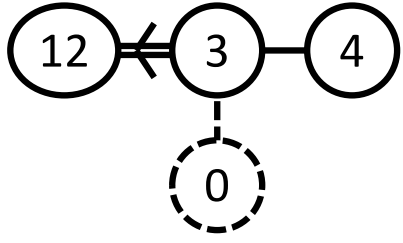}
\caption{The $\widehat B_3$ Dynkin diagram.}
\label{affineB3}
\end{figure}

It is interesting to note that the inequality \eqref{boundaryB3} can be written as
\begin{align}
(\theta,\lambda)\le k,
\label{affineineq}
\end{align}
(compare with (14.127) in \cite{CFT}) where
\begin{align}
\theta=(-1,1/2,-1/2)=2\alpha_{12}+2\alpha_3+\alpha_4,
\end{align}
is the highest root of the $B_3$ algebra.
In other words, by introducing the finite part of the $\widehat B_3$ affine root $\alpha_0=-\theta$, we can express \eqref{boundaryB3} as
\begin{align}
(\alpha_0,\lambda)+k\ge 0,
\end{align}
or even simply
\begin{align}
(\widehat\alpha_i,\widehat\lambda)\ge 0\;\;\;(i=0,12,3,4),
\end{align}
including the condition of the $B_3$ Weyl chamber \eqref{B3chamber}, if we introduce the affine weight space $\widehat\lambda=(\lambda,k_\lambda,n_\lambda)$ (with the level $k$ and the grading $n$) and consider the inner product $(\widehat\lambda,\widehat\mu)=(\lambda,\mu)+k_\lambda n_\mu+k_\mu n_\lambda$ with finite simple roots $\widehat\alpha_i=(\alpha_i,0,0)$ and the affine simple root $\widehat\alpha_0=(\alpha_0,0,1)$ (see (14.23) and (14.32) in \cite{CFT}). 
Then, the extended Cartan matrix
\begin{align}
\widehat A_{ij}=\frac{2(\widehat\alpha_{i},\widehat\alpha_{j})}{(\widehat\alpha_{j},\widehat\alpha_{j})},
\label{cartan}
\end{align}
is given by
\begin{align}
\widehat A=\begin{pmatrix}2&-1&0&0\\-2&2&-1&-1\\0&-1&2&0\\0&-1&0&2\end{pmatrix},
\end{align}
with $i,j=12,3,4,0$.
This is nothing but the extended Cartan matrix for the affine algebra of $B_3$, which we denote by $\widehat B_3$ hereafter.
See figure \ref{affineB3} for the $\widehat B_3$ Dynkin diagram.

\begin{figure}[!t]
\centering\includegraphics[scale=0.6,angle=0]{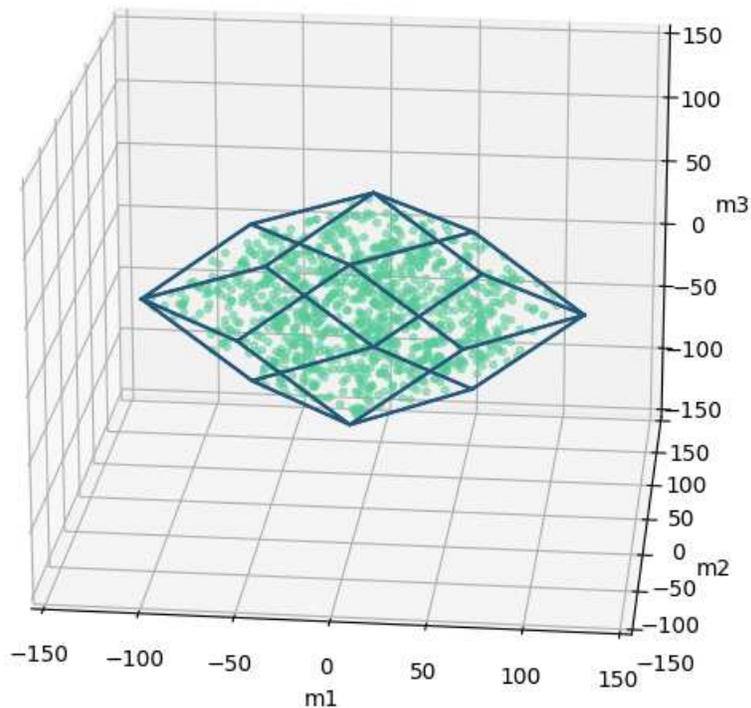}
\caption{Numerical analysis of duality cascades.
The green dots denote the endpoints of duality cascades obtained from various initial brane configurations and the blue lines form the fundamental domain of the rhombic dodecahedron studied in figure \ref{domainB3}.}
\label{numerical}
\end{figure}

We have also performed numerical analysis to study the final destinations of duality cascades.
By preparing various initial brane configurations and applying various processes of duality cascades using the HW transitions, finally we obtain a set of brane configurations.
We depict the final destinations in figure \ref{numerical}.
It is clear that this matches the fundamental domain of our rhombic dodecahedron.

To summarize, here we start simply from brane configurations with two NS5-branes and two $(1,k)$5-branes and study the fundamental domain of duality cascades by requiring that the reference rank continues to be the lowest in applications of the HW transitions.
Since the system of brane configurations is embedded in the $D_5$ quantum curve, it may not be surprising that the condition enjoys symmetries of the $B_3$ Weyl group as the invariant subgroup.
It is, however, rather surprising to us to encounter the affine $\widehat B_3$ Weyl group here.
Namely, after constraining the fundamental domain into the $B_3$ Weyl chamber, what we find is the affine $\widehat B_3$ Weyl chamber.

\subsubsection{Duality cascades ($B_3$)}

So far we have studied the fundamental domain with the reference fixed.
We expect this to consist of all the final destinations of duality cascades for supersymmetric brane configurations.
Now let us turn to the duality cascades themselves.
It is proposed that, in terms of brane configurations, the duality cascades are equivalent to the following processes.
\begin{enumerate}
\item Consider a brane configuration with one of the lowest ranks labeled as the reference.
\item When we encounter lower ranks by applying the HW transitions arbitrarily, relabel one of the lowest ranks as the reference.
\item Repeat the above step until lower ranks do not appear any more. 
\end{enumerate}
It is, however, unclear whether the processes always end and whether the endpoint is unique depending only on the initial brane configuration.
Here let us start by studying the step of changing references.

\begin{table}[t!]
\begin{center}
\begin{tabular}{c||c|c}
$\langle\stackrel{2}{\bullet}\stackrel{1}{\bullet}\stackrel{3}{\circ}\stackrel{4}{\circ}\rangle$&$(M_1,M_2,M_3)$&$N=N_0+M_2+M_3$\\\hline
$\langle\stackrel{2}{\bullet}\cdot\stackrel{1}{\bullet}\stackrel{3}{\circ}\stackrel{4}{\circ}\rangle$&$(M_1+k,M_2-\frac{k}{2},M_3+\frac{k}{2})$&$N+M_1-M_2+M_3+k$\\
$\langle\stackrel{1}{\bullet}\cdot\stackrel{2}{\bullet}\stackrel{3}{\circ}\stackrel{4}{\circ}\rangle$&$(M_1+k,M_2+\frac{k}{2},M_3-\frac{k}{2})$&$N+M_1+M_2-M_3+k$\\
$\langle\stackrel{3}{\circ}\cdot\stackrel{2}{\bullet}\stackrel{1}{\bullet}\stackrel{4}{\circ}\rangle$&$(M_1-k,M_2-\frac{k}{2},M_3-\frac{k}{2})$&$N-M_1-M_2-M_3+k$\\
$\langle\stackrel{4}{\circ}\cdot\stackrel{2}{\bullet}\stackrel{1}{\bullet}\stackrel{3}{\circ}\rangle$&$(M_1-k,M_2+\frac{k}{2},M_3+\frac{k}{2})$&$N-M_1+M_2+M_3+k$\\
$\langle\stackrel{2}{\bullet}\stackrel{1}{\bullet}\cdot\stackrel{3}{\circ}\stackrel{4}{\circ}\rangle$&$(M_1+2k,M_2,M_3)$&$N+2M_1+2k$\\
$\langle\stackrel{2}{\bullet}\stackrel{3}{\circ}\cdot\stackrel{1}{\bullet}\stackrel{4}{\circ}\rangle$&$(M_1,M_2-k,M_3)$&$N-2M_2+k$\\
$\langle\stackrel{1}{\bullet}\stackrel{3}{\circ}\cdot\stackrel{2}{\bullet}\stackrel{4}{\circ}\rangle$&$(M_1,M_2,M_3-k)$&$N-2M_3+k$\\
$\langle\stackrel{2}{\bullet}\stackrel{4}{\circ}\cdot\stackrel{1}{\bullet}\stackrel{3}{\circ}\rangle$&$(M_1,M_2,M_3+k)$&$N+2M_3+k$\\
$\langle\stackrel{1}{\bullet}\stackrel{4}{\circ}\cdot\stackrel{2}{\bullet}\stackrel{3}{\circ}\rangle$&$(M_1,M_2+k,M_3)$&$N+2M_2+k$\\
$\langle\stackrel{3}{\circ}\stackrel{4}{\circ}\cdot\stackrel{2}{\bullet}\stackrel{1}{\bullet}\rangle$&$(M_1-2k,M_2,M_3)$&$N-2M_1+2k$\\
$\langle\stackrel{2}{\bullet}\stackrel{1}{\bullet}\stackrel{3}{\circ}\cdot\stackrel{4}{\circ}\rangle$&$(M_1+k,M_2-\frac{k}{2},M_3-\frac{k}{2})$&$N+M_1-M_2-M_3+k$\\
$\langle\stackrel{2}{\bullet}\stackrel{1}{\bullet}\stackrel{4}{\circ}\cdot\stackrel{3}{\circ}\rangle$&$(M_1+k,M_2+\frac{k}{2},M_3+\frac{k}{2})$&$N+M_1+M_2+M_3+k$\\
$\langle\stackrel{2}{\bullet}\stackrel{3}{\circ}\stackrel{4}{\circ}\cdot\stackrel{1}{\bullet}\rangle$&$(M_1-k,M_2-\frac{k}{2},M_3+\frac{k}{2})$&$N-M_1-M_2+M_3+k$\\
$\langle\stackrel{1}{\bullet}\stackrel{3}{\circ}\stackrel{4}{\circ}\cdot\stackrel{2}{\bullet}\rangle$&$(M_1-k,M_2+\frac{k}{2},M_3-\frac{k}{2})$&$N-M_1+M_2-M_3+k$
\end{tabular}
\end{center}
\caption{Changes of the relative ranks $(M_1,M_2,M_3)$ and the overall rank $N=N_0+M_2+M_3$, for the brane configurations of the $B_3$ Weyl group.
Here we change references into that denoted by $\cdot$ in applying duality cascades.
Note that, instead of the original notation of the overall rank $N_0$, we consider the reference rank $N=N_0+M_2+M_3$.}
\label{changerefB3}
\end{table}

In \cite{KM} the cyclic changes of references were already studied.
It was found that, in the parameter space $\{(M_1,M_2,M_3)\}$, changing references only induces translational shifts.
This is natural since the HW transitions can be understood from charge conservation \eqref{charge} of the difference of the numbers of 5-branes and adjacent D3-branes located on both sides.

In this paper we present a full study of changes of references appearing in the HW transitions.
It seems that the above reasoning of translations remains applicable.
We study the translations for each change of references and list the results in table \ref{changerefB3}.
Namely, we apply various brane transitions to the brane configuration $\langle\stackrel{2}{\bullet}\stackrel{1}{\bullet}\stackrel{3}{\circ}\stackrel{4}{\circ}\rangle$ and change references to those denoted by $\cdot$.
Then, by returning to the original order, the parameters are found to be shifted as in the table.
For example, for the case of $\langle\stackrel{2}{\bullet}\stackrel{3}{\circ}\cdot\stackrel{1}{\bullet}\stackrel{4}{\circ}\rangle$, we first apply the HW transition to $\stackrel{3}{\circ}$ and $\stackrel{1}{\bullet}$ for the original brane configuration as in \eqref{HWex} and then change references as
\begin{align}
&\langle N_0-M_2+M_3+k\stackrel{1}{\bullet}N_0+M_1+k\stackrel{4}{\circ}N_0+M_2+M_3\stackrel{2}{\bullet}N_0+M_1+2M_3+k\stackrel{3}{\circ}\rangle\nonumber\\
&=\langle N_0-M_2+M_3+k\stackrel{2}{\bullet}N_0+M_1-2M_2+2M_3+3k\stackrel{1}{\bullet}\nonumber\\
&\qquad N_0+2M_1-M_2+M_3+3k\stackrel{3}{\circ}N_0+M_1-2M_2+3k\stackrel{4}{\circ}\rangle.
\end{align}
By comparing with the original brane configuration \eqref{M123}, we find the variables are changed as in table \ref{changerefB3}.

We can present a rather clear interpretation for the results in table \ref{changerefB3}, by comparing it with table \ref{inequalityB3}.
Namely, in table \ref{inequalityB3} we have moved into various orders of 5-branes and found inequalities imposed by comparing various ranks with the reference.
Geometrically, this defines faces of the rhombic dodecahedron.
If an inequality does not hold, or in other words, lower ranks appear in some intervals, we change references in duality cascades, which causes parameter shifts given in table \ref{changerefB3}.
Geometrically, this shifts the parameters of the relative ranks perpendicularly to the face whose inequality fails, so that the failure of the inequality can be improved, and decreases the overall rank $N$ accordingly.
For example, if the inequality $2M_2\le k$ imposed from \eqref{HWex} fails and we change references into $\cdot$ in $\langle\stackrel{2}{\bullet}\stackrel{3}{\circ}\cdot\stackrel{1}{\bullet}\stackrel{4}{\circ}\rangle$, parameters of the relative ranks $(M_1,M_2,M_3)$ are shifted perpendicularly to $(M_1,M_2-k,M_3)$ so that the failure is improved and the overall rank $N$ decreases to $N-2M_2+k$ accordingly.

Now let us come back to our original questions raised in the introduction.
\begin{itemize}
\item
Starting from an arbitrary brane configuration, by repeating duality cascades, do we always arrive at a final brane configuration, which cannot be changed by duality cascades any more?
\item
Is the final destination of duality cascades uniquely determined from the initial brane configuration?
\end{itemize}
Since duality cascades are realized by the translations in the parameter space, we can rephrase the above questions in terms of the geometrical property of the fundamental domain as follows.
\begin{itemize}
\item
Starting at an arbitrary point $(M_1,M_2,M_3)$, by subsequent applications of the translations, do we finally arrive in the fundamental domain?
Conversely, by subsequent applications of the translations, does the fundamental domain cover the whole parameter space?
\item
Is the final destination of the translations in the fundamental domain uniquely determined from the initial parameters?
Conversely, by subsequent applications of the translations, do all the copies of the fundamental domain either coincide or be disjoint sharing at most their boundaries?
\end{itemize}
In short, the question is whether the fundamental domain can fill the whole parameter space by translations.
This is nothing but the idea of {\it parallelotopes}.
Hence, for the finiteness of the processes and the uniqueness of the endpoint of duality cascades, we propose that the fundamental domains of duality cascades should be parallelotopes.

The property of the affine Weyl group suggested in \eqref{affineineq} helps to understand the relation between fundamental domains of duality cascades and parallelotopes.
For the $\widehat B_3$ case, we observe that the transformation of the overall rank $N$ for $\langle\stackrel{2}{\bullet}\cdot\stackrel{1}{\bullet}\stackrel{3}{\circ}\stackrel{4}{\circ}\rangle$ in table \ref{changerefB3} can be rewritten as
\begin{align}
N\to N-(\theta,\lambda)+k.
\label{Nreflection}
\end{align}
Interestingly, this coincides with the affine part of the reflection $s_0$, where the overall rank $N$ plays the role of the eigenvalue of the grading operator $L_0=-t(d/dt)$ in representations of the affine Weyl group (compare with (14.57) (14.73) in \cite{CFT}).
Then, the translation in the affine Weyl group
\begin{align}
t_0=s_0s_3s_{12}s_3s_4s_3s_{12}s_3,
\end{align}
is identified as the change of references into $\langle\stackrel{2}{\bullet}\cdot\stackrel{1}{\bullet}\stackrel{3}{\circ}\stackrel{4}{\circ}\rangle$ in duality cascades with the finite part of $s_0$ being
\begin{align}
s_0:(M_1,M_2,M_3)\mapsto\bigg(M_2-M_3-k,\frac{M_1+M_2+M_3+k}{2},\frac{-M_1+M_2+M_3-k}{2}\bigg),
\label{s0M123}
\end{align}
which reflects with respect to the plane $-M_1+M_2-M_3=k$ preserving the points \eqref{convexB3}.
Thus, by combining the translation $t_0$ with the Weyl group, we obtain the whole affine Weyl group containing $s_0$.
After equipped with the affine Weyl reflection $s_0$, translations in other directions (corresponding to changes into other references) can be generated by various elements in the affine Weyl group.

Since the translations $t$ are elements of the affine Weyl group, they do not change the norm $(\widehat{\lambda},\widehat{\lambda})$; the norms for $\widehat{\lambda}=(\lambda,k,-N)$ and $t\widehat{\lambda}=(\lambda',k,-N')$ are equal,
\begin{align}
(\widehat{\lambda},\widehat{\lambda})=(t\widehat{\lambda},t\widehat{\lambda}),
\end{align}
or, by using $(\widehat{\lambda},\widehat{\mu})=(\lambda,\mu)+k_\lambda n_\nu+k_\mu n_\lambda$,
\begin{align}
(\lambda',\lambda')-(\lambda,\lambda)=2k(N'-N).
\end{align}
Since the duality cascades decrease the overall rank $N$, this implies that the duality cascades decrease the norm $(\lambda,\lambda)=M_1^2+2M_2^2+2M_3^2$ as well.
This argument gives us a geometrically intuitive picture how the duality cascades end by falling down to the fundamental domain around the origin in the space of the relative ranks $\{\lambda=(M_1,M_2,M_3)\}$.

In summary, following the previous discovery of the affine $\widehat B_3$ Weyl chamber for the fundamental domain, here we can proceed further.
We have explained that the duality cascades can be realized by translations and identified the translations as those constructed from the affine Weyl group.
The affine Weyl group guarantees that the fundamental domain forms a parallelotope, which further ensures the physical properties that duality cascades always end in finite processes and that the endpoint of duality cascades is unique depending only on the initial brane configuration.

\subsection{$F_4$}\label{sectionf4}

Now let us turn to the brane configurations corresponding to the $E_7$ quantum curve.
This time the brane configurations consist of two NS5-branes and four $(1,k)$5-branes (placed standardly in the order of $\langle{\bullet}{\circ}{\circ}{\bullet}{\circ}{\circ}\rangle$) with a constraint for the relative ranks and the invariant subgroup is the $F_4$ Weyl group (see appendix \ref{appendixE7}).
The corresponding gauge theories described by these brane configurations have gauge group $\text{U}(N_1)_k\times\text{U}(N_2)_{-k}\times\text{U}(N_3)_0\times\text{U}(N_4)_k\times\text{U}(N_5)_{-k}\times\text{U}(N_6)_0$ and pairs of bifundamental matters connecting adjacent factors of the gauge group cyclically, where the ranks and the levels are subject to change due to the HW transitions.

\subsubsection{Fundamental domain ($F_4$)}

As in \cite{FMN} we label the relative ranks of the brane configurations by
\begin{align}
&\langle N_1\stackrel{\text{ii}}{\bullet}
N_2\stackrel{1}{\circ}
N_3\stackrel{2}{\circ}
N_4\stackrel{\text{i}}{\bullet}
N_5\stackrel{3}{\circ}
N_6\stackrel{4}{\circ}\rangle
=\Big\langle N\stackrel{\text{ii}}{\bullet}
N+G_1+k\stackrel{1}{\circ}
N+\frac{1}{2}(-3F_1+F_2+F_3+G_1)+k\stackrel{2}{\circ}\nonumber\\
&N-F_1-F_2+F_3+k\stackrel{\text{i}}{\bullet}
N-F_1-F_2+F_3+G_1+2k\stackrel{3}{\circ}
N+\frac{1}{2}(-F_1-F_2-F_3+G_1)+k\stackrel{4}{\circ}\Big\rangle,
\label{F4braneconfig}
\end{align}
so that the relative ranks $(F_1,F_2,F_3,G_1)$ match the coefficients $(f_1,f_2,f_3,g_1)$ for the $E_7$ quantum curve \eqref{F4subspace} (after the ${\mathbb Z}_2$ folding is applied) by the identification $(f_1,f_2,f_3,g_1)=(e^{2\pi iF_1},e^{2\pi iF_2},e^{2\pi iF_3},e^{2\pi iG_1})$.
Note that, for symmetries of the Weyl group to work, we need to constrain the parameters by the ``balanced'' condition \cite{FMN}
\begin{align}
N_1+N_5=N_2+N_4.
\label{balance}
\end{align}

Let us define the fundamental domain of duality cascades by requiring the same condition (as the previous $B_3$ case) that the reference rank $N_1=N$ continues to be the lowest rank even we apply various brane transitions.
Compared with the $B_3$ case, the condition should be studied carefully due to the constraint \eqref{balance}.
For this reason, let us avoid arbitrary applications of the HW transitions and consider only the brane configurations with 5-branes in the orders of ${\langle}{\stackrel{\text{ii}}{\bullet}}{\circ}{\circ}{\stackrel{\text{i}}{\bullet}}{\circ}{\circ}{\rangle}$, ${\langle}{\circ}{\stackrel{\text{ii}}{\bullet}}{\circ}{\circ}{\stackrel{\text{i}}{\bullet}}{\circ}{\rangle}$ and ${\langle}{\circ}{\circ}{\stackrel{\text{ii}}{\bullet}}{\circ}{\circ}{\stackrel{\text{i}}{\bullet}}{\rangle}$.
Then, we find 24 inequalities
\begin{align}
&|F_1+F_2-F_3|\le k,\quad
|F_1-F_2+F_3|\le k,\quad
|{-F_1+F_2+F_3}|\le k,\nonumber\\
&|G_1|\le k,\quad|F_1+F_2+F_3\pm G_1|\le 2k,\nonumber\\
&|{-3F_1+F_2+F_3\pm G_1}|\le 2k,\quad
|F_1-3F_2+F_3\pm G_1|\le 2k,\quad
|F_1+F_2-3F_3\pm G_1|\le 2k.
\label{domainF4ineq}
\end{align}
Note that inequalities such as $|\pm F_1\pm F_2\pm F_3+G_1|\le 2k$ also appear, though these are trivially guaranteed from the others.
Also, even if we consider brane configurations in arbitrary orders of 5-branes in addition (with only the order of the two NS5-branes fixed) to study the requirements of inequalities, extra inequalities are all guaranteed.

Compared with the three-dimensional rhombic dodecahedron, this is a four-dimensional polytope and difficult to visualize.
However, with the viewpoint of Weyl groups, we can analyze the fundamental domain with the same technique.
As in the $B_3$ case, using the $F_4$ Weyl group in appendix \ref{appendixE7}, we can restrict the analysis of the fundamental domain to the $F_4$ Weyl chamber
\begin{align}
&(\alpha_{16},\lambda)=F_2-F_3\ge 0,\quad(\alpha_{25},\lambda)=F_1-F_2\ge 0,\nonumber\\
&(\alpha_3,\lambda)=(-3F_1+F_2+F_3-G_1)/2\ge 0,\quad(\alpha_4,\lambda)=G_1\ge 0,
\end{align}
with the metric for the bilinear form given in \eqref{F4metric}, the simple roots $\alpha_{16}$, $\alpha_{25}$, $\alpha_3$ and $\alpha_4$ in \eqref{F4sr} and $\lambda=(F_1,F_2,F_3,G_1)$.
This time, in the four-dimensional space, the edges of the $F_4$ Weyl chamber is fixed by three hyperplanes.
By solving for intersections of the fundamental domain along these edges, we find
\begin{align}
k\omega_{16},\quad
k\omega_{25}/2,\quad
k\omega_3/3,\quad
k\omega_4/2,
\label{pointsF4}
\end{align}
where $\omega_{16}$, $\omega_{25}$, $\omega_3$ and $\omega_4$ are the fundamental weights \eqref{F4fw} and the inverse comarks appear in the coefficients.
Hence, the fundamental domain in the $F_4$ Weyl chamber has to contain the convex hull of the above points and the origin.
Since the above points lie on the hyperplane $-F_1-F_2-F_3+G_1=2k$ and the inequality
\begin{align}
-F_1-F_2-F_3+G_1\le 2k,
\label{ineqF4}
\end{align}
already appears in the condition for the fundamental domain \eqref{domainF4ineq}, the fundamental domain in the $F_4$ Weyl chamber is exactly the four-dimensional subspace surrounded by \eqref{ineqF4} and these edges.

\begin{figure}[!t]
\centering\includegraphics[scale=0.7,angle=-90]{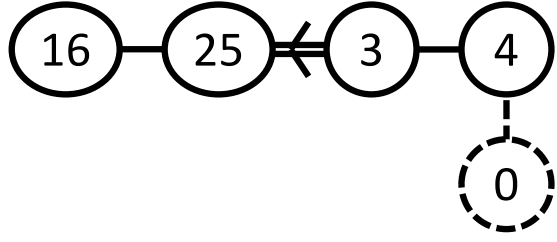}
\caption{The $\widehat F_4$ Dynkin diagram.}
\label{affineF4}
\end{figure}

Interestingly, note that \eqref{ineqF4} can be rewritten as \eqref{affineineq} again, where
\begin{align}
\theta=(-1,-1,-1,1)=2\alpha_{16}+4\alpha_{25}+3\alpha_3+2\alpha_4,
\end{align}
is the highest root of the $F_4$ algebra.
Namely, as in the $B_3$ case, the fundamental domain of duality cascades in the $F_4$ Weyl chamber is nothing but the affine $\widehat F_4$ Weyl chamber.
The $\widehat F_4$ Dynkin diagram is depicted in figure \ref{affineF4} and the extended Cartan matrix \eqref{cartan} is given by
\begin{align}
\widehat A=\begin{pmatrix}
2&-1&0&0&0\\
-1&2&-1&0&0\\
0&-2&2&-1&0\\
0&0&-1&2&-1\\
0&0&0&-1&2
\end{pmatrix},
\end{align}
for $i,j=16,25,3,4,0$ with $0$ denoting the $\widehat F_4$ affine root.

\subsubsection{Duality cascades ($F_4$)}

\begin{figure}[!t]
\centering\includegraphics[scale=0.6,angle=-90]{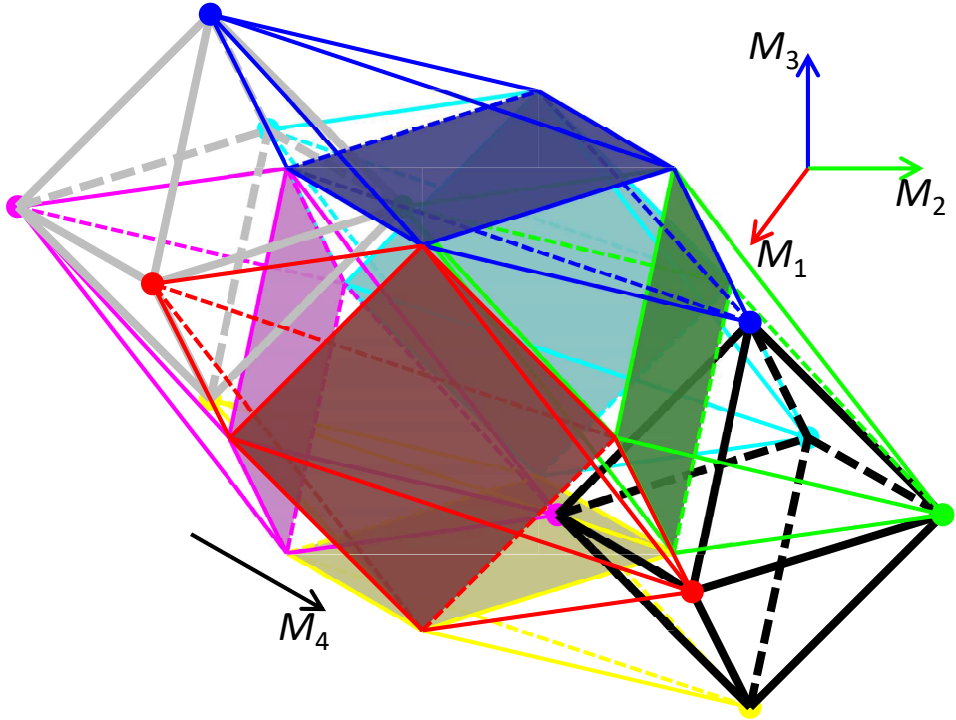}
\caption{Four-dimensional 24-cell consisting of 24 cells, 96 edges and 24 vertices.
We specialize the direction of $M_4$.
The vertices in $M_4=\pm k$ form the octahedra, while the vertices in $M_4=0$ form the cuboctahedron.
We depict the octahedra in black fat lines, while the cuboctahedron with color faces.
Differences between solid and dashed lines do not make much sense visually (since it is four-dimensional).
Aside from those in $M_4=\pm k$, the remaining 22 boundary octahedra are given by the following two types.
6 of them connecting the corresponding vertices and squares spread among the three polyhedra (depicted in various colors), while 16 of them connecting the corresponding triangular faces in two adjacent polyhedra (with each 8 connecting faces of the central cuboctahedron and each octahedron, not depicted to avoid confusion).}
\label{24cellfig}
\end{figure}

As in the $B_3$ case, it is interesting to ask whether, after the duality cascades, the brane configuration always reduces to that in the fundamental domain and whether the final brane configuration is unique, which is geometrically equivalent to whether the fundamental domain forms a parallelotope.
Again, this should be the case by following all of our previous discussions of the affine Weyl chamber.

It is, however, interesting to see that the fundamental domain of duality cascades forms a parallelotope more directly.
For this purpose we first notice that, as in the $B_3$ case, the vertices of the polytope consist of the brane configurations with various orders of 5-branes (without rank differences).
We can then study various geometrical properties from the definition of the boundary hyperplanes in \eqref{domainF4ineq} and the 24 vertices listed in \cite{FMN} (and repeated below in table \ref{24cell}).
For example, it is easy to find that all hyperplanes contain 6 vertices and all vertices reside on 6 hyperplanes simultaneously.
Interestingly, all of these geometrical properties suggest that the fundamental domain is the famous 24-cell (see figure \ref{24cellfig}).
The boundary of the 24-cell consists of 24 octahedra, which share their boundary triangles with the adjacent ones.
Importantly, the 24-cell can tile the whole four-dimensional space by translations, hence is a parallelotope.

\begin{table}[t!]
\begin{center}
\begin{tabular}{l||l|l|l}
type&brane config.&$(F_1,F_2,F_3,G_1)$&$(M_1,M_2,M_3,M_4)$\\\hline\hline
$\langle\stackrel{\text{ii}}{\bullet}\stackrel{}{\circ}\stackrel{}{\circ}\stackrel{\text{i}}{\bullet}\stackrel{}{\circ}\stackrel{}{\circ}\rangle$
&$\langle\stackrel{\text{ii}}{\bullet}\,\stackrel{1}{\circ}\,\stackrel{2}{\circ}\,\stackrel{\text{i}}{\bullet}\,\stackrel{3}{\circ}\,\stackrel{4}{\circ}\rangle$
&$(\frac{k}{2},\frac{k}{2},0,-k)$&$(0,0,-k,-k)$\\\cline{2-4}
&$\langle\stackrel{\text{ii}}{\bullet}\,\stackrel{1}{\circ}\,\stackrel{3}{\circ}\,\stackrel{\text{i}}{\bullet}\,\stackrel{2}{\circ}\,\stackrel{4}{\circ}\rangle$
&$(\frac{k}{2},0,\frac{k}{2},-k)$&$(0,-k,0,-k)$\\\cline{2-4}
&$\langle\stackrel{\text{ii}}{\bullet}\,\stackrel{2}{\circ}\,\stackrel{3}{\circ}\,\stackrel{\text{i}}{\bullet}\,\stackrel{1}{\circ}\,\stackrel{4}{\circ}\rangle$
&$(0,\frac{k}{2},\frac{k}{2},-k)$&$(-k,0,0,-k)$\\\cline{2-4}
&$\langle\stackrel{\text{ii}}{\bullet}\,\stackrel{1}{\circ}\,\stackrel{4}{\circ}\,\stackrel{\text{i}}{\bullet}\,\stackrel{2}{\circ}\,\stackrel{3}{\circ}\rangle$
&$(0,-\frac{k}{2},-\frac{k}{2},-k)$&$(k,0,0,-k)$\\\cline{2-4}
&$\langle\stackrel{\text{ii}}{\bullet}\,\stackrel{2}{\circ}\,\stackrel{4}{\circ}\,\stackrel{\text{i}}{\bullet}\,\stackrel{1}{\circ}\,\stackrel{3}{\circ}\rangle$
&$(-\frac{k}{2},0,-\frac{k}{2},-k)$&$(0,k,0,-k)$\\\cline{2-4}
&$\langle\stackrel{\text{ii}}{\bullet}\,\stackrel{3}{\circ}\,\stackrel{4}{\circ}\,\stackrel{\text{i}}{\bullet}\,\stackrel{1}{\circ}\,\stackrel{2}{\circ}\rangle$
&$(-\frac{k}{2},-\frac{k}{2},0,-k)$&$(0,0,k,-k)$\\\hline\hline
$\langle\stackrel{}{\circ}\stackrel{\text{ii}}{\bullet}\stackrel{}{\circ}\stackrel{}{\circ}\stackrel{\text{i}}{\bullet}\stackrel{}{\circ}\rangle$
&$\langle\stackrel{1}{\circ}\,\stackrel{\text{ii}}{\bullet}\,\stackrel{2}{\circ}\,\stackrel{3}{\circ}\,\stackrel{\text{i}}{\bullet}\,\stackrel{4}{\circ}\rangle$
&$(k,\frac{k}{2},\frac{k}{2},0)$&$(0,-k,-k,0)$\\\cline{2-4}
&$\langle\stackrel{2}{\circ}\,\stackrel{\text{ii}}{\bullet}\,\stackrel{1}{\circ}\,\stackrel{3}{\circ}\,\stackrel{\text{i}}{\bullet}\,\stackrel{4}{\circ}\rangle$
&$(\frac{k}{2},k,\frac{k}{2},0)$&$(-k,0,-k,0)$\\\cline{2-4}
&$\langle\stackrel{3}{\circ}\,\stackrel{\text{ii}}{\bullet}\,\stackrel{1}{\circ}\,\stackrel{2}{\circ}\,\stackrel{\text{i}}{\bullet}\,\stackrel{4}{\circ}\rangle$
&$(\frac{k}{2},\frac{k}{2},k,0)$&$(-k,-k,0,0)$\\\cline{2-4}
&$\langle\stackrel{1}{\circ}\,\stackrel{\text{ii}}{\bullet}\,\stackrel{2}{\circ}\,\stackrel{4}{\circ}\,\stackrel{\text{i}}{\bullet}\,\stackrel{3}{\circ}\rangle$
&$(\frac{k}{2},0,-\frac{k}{2},0)$&$(k,0,-k,0)$\\\cline{2-4}
&$\langle\stackrel{2}{\circ}\,\stackrel{\text{ii}}{\bullet}\,\stackrel{1}{\circ}\,\stackrel{4}{\circ}\,\stackrel{\text{i}}{\bullet}\,\stackrel{3}{\circ}\rangle$
&$(0,\frac{k}{2},-\frac{k}{2},0)$&$(0,k,-k,0)$\\\cline{2-4}
&$\langle\stackrel{4}{\circ}\,\stackrel{\text{ii}}{\bullet}\,\stackrel{1}{\circ}\,\stackrel{2}{\circ}\,\stackrel{\text{i}}{\bullet}\,\stackrel{3}{\circ}\rangle$
&$(-\frac{k}{2},-\frac{k}{2},-k,0)$&$(k,k,0,0)$\\\cline{2-4}
&$\langle\stackrel{1}{\circ}\,\stackrel{\text{ii}}{\bullet}\,\stackrel{3}{\circ}\,\stackrel{4}{\circ}\,\stackrel{\text{i}}{\bullet}\,\stackrel{2}{\circ}\rangle$
&$(\frac{k}{2},-\frac{k}{2},0,0)$&$(k,-k,0,0)$\\\cline{2-4}
&$\langle\stackrel{3}{\circ}\,\stackrel{\text{ii}}{\bullet}\,\stackrel{1}{\circ}\,\stackrel{4}{\circ}\,\stackrel{\text{i}}{\bullet}\,\stackrel{2}{\circ}\rangle$
&$(0,-\frac{k}{2},\frac{k}{2},0)$&$(0,-k,k,0)$\\\cline{2-4}
&$\langle\stackrel{4}{\circ}\,\stackrel{\text{ii}}{\bullet}\,\stackrel{1}{\circ}\,\stackrel{3}{\circ}\,\stackrel{\text{i}}{\bullet}\,\stackrel{2}{\circ}\rangle$
&$(-\frac{k}{2},-k,-\frac{k}{2},0)$&$(k,0,k,0)$\\\cline{2-4}
&$\langle\stackrel{2}{\circ}\,\stackrel{\text{ii}}{\bullet}\,\stackrel{3}{\circ}\,\stackrel{4}{\circ}\,\stackrel{\text{i}}{\bullet}\,\stackrel{1}{\circ}\rangle$
&$(-\frac{k}{2},\frac{k}{2},0,0)$&$(-k,k,0,0)$\\\cline{2-4}
&$\langle\stackrel{3}{\circ}\,\stackrel{\text{ii}}{\bullet}\,\stackrel{2}{\circ}\,\stackrel{4}{\circ}\,\stackrel{\text{i}}{\bullet}\,\stackrel{1}{\circ}\rangle$
&$(-\frac{k}{2},0,\frac{k}{2},0)$&$(-k,0,k,0)$\\\cline{2-4}
&$\langle\stackrel{4}{\circ}\,\stackrel{\text{ii}}{\bullet}\,\stackrel{2}{\circ}\,\stackrel{3}{\circ}\,\stackrel{\text{i}}{\bullet}\,\stackrel{1}{\circ}\rangle$
&$(-k,-\frac{k}{2},-\frac{k}{2},0)$&$(0,k,k,0)$\\\hline\hline
$\langle\stackrel{}{\circ}\stackrel{}{\circ}\stackrel{\text{ii}}{\bullet}\stackrel{}{\circ}\stackrel{}{\circ}\stackrel{\text{i}}{\bullet}\rangle$
&$\langle\stackrel{1}{\circ}\,\stackrel{2}{\circ}\,\stackrel{\text{ii}}{\bullet}\,\stackrel{3}{\circ}\,\stackrel{4}{\circ}\,\stackrel{\text{i}}{\bullet}\rangle$
&$(\frac{k}{2},\frac{k}{2},0,k)$&$(0,0,-k,k)$\\\cline{2-4}
&$\langle\stackrel{1}{\circ}\,\stackrel{3}{\circ}\,\stackrel{\text{ii}}{\bullet}\,\stackrel{2}{\circ}\,\stackrel{4}{\circ}\,\stackrel{\text{i}}{\bullet}\rangle$
&$(\frac{k}{2},0,\frac{k}{2},k)$&$(0,-k,0,k)$\\\cline{2-4}
&$\langle\stackrel{2}{\circ}\,\stackrel{3}{\circ}\,\stackrel{\text{ii}}{\bullet}\,\stackrel{1}{\circ}\,\stackrel{4}{\circ}\,\stackrel{\text{i}}{\bullet}\rangle$
&$(0,\frac{k}{2},\frac{k}{2},k)$&$(-k,0,0,k)$\\\cline{2-4}
&$\langle\stackrel{1}{\circ}\,\stackrel{4}{\circ}\,\stackrel{\text{ii}}{\bullet}\,\stackrel{2}{\circ}\,\stackrel{3}{\circ}\,\stackrel{\text{i}}{\bullet}\rangle$
&$(0,-\frac{k}{2},-\frac{k}{2},k)$&$(k,0,0,k)$\\\cline{2-4}
&$\langle\stackrel{2}{\circ}\,\stackrel{4}{\circ}\,\stackrel{\text{ii}}{\bullet}\,\stackrel{1}{\circ}\,\stackrel{3}{\circ}\,\stackrel{\text{i}}{\bullet}\rangle$
&$(-\frac{k}{2},0,-\frac{k}{2},k)$&$(0,k,0,k)$\\\cline{2-4}
&$\langle\stackrel{3}{\circ}\,\stackrel{4}{\circ}\,\stackrel{\text{ii}}{\bullet}\,\stackrel{1}{\circ}\,\stackrel{2}{\circ}\,\stackrel{\text{i}}{\bullet}\rangle$
&$(-\frac{k}{2},-\frac{k}{2},0,k)$&$(0,0,k,k)$
\end{tabular}
\end{center}
\caption{The parameters $(F_1,F_2,F_3,G_1)$ and $(M_1,M_2,M_3,M_4)$ for the brane configurations with various orders of 5-branes (without rank differences).}
\label{24cell}
\end{table}

Since it is known that the vertices of the 24-cell is described by the 24 permutations of $(\pm 1,\pm 1,0,0)$ (with any combination of multiple signs), let us make a variable change
\begin{align}
M_1=F_1-F_2-F_3,\quad
M_2=-F_1+F_2-F_3,\quad
M_3=-F_1-F_2+F_3,\quad
M_4=G_1.
\label{M1234}
\end{align}
In table \ref{24cell} we list the original parameters $(F_1,F_2,F_3,G_1)$ and the new ones $(M_1,M_2,M_3,M_4)$ for the brane configurations without rank differences.
Also, the hyperplanes \eqref{domainF4ineq} are rewritten simply as
\begin{align}
\pm M_1\le k,\quad\pm M_2\le k,\quad\pm M_3\le k,\quad\pm M_4\le k,\quad
\pm M_1\pm M_2\pm M_3\pm M_4\le 2k,
\label{24cellineq}
\end{align}
in terms of our new variables \eqref{M1234}, where any combination of multiple signs may apply.
The four-dimensional 24-cell in figure \ref{24cellfig} is depicted with these new variables.

\begin{table}[p!]
\begin{center}
\begin{tabular}{c||c|c}
$\langle\stackrel{\text{ii}}{\bullet}\stackrel{1}{\circ}\stackrel{2}{\circ}\stackrel{\text{i}}{\bullet}\stackrel{3}{\circ}\stackrel{4}{\circ}\rangle$&$(M_1,M_2,M_3,M_4)$&$N$\\\hline\hline
$\langle\stackrel{\text{ii}}{\bullet}\stackrel{\text{i}}{\bullet}\stackrel{1}{\circ}\stackrel{2}{\circ}\stackrel{3}{\circ}\cdot\stackrel{4}{\circ}\rangle$&$(M_1+k,M_2+k,M_3+k,M_4+k)$&$N+(M_1+M_2+M_3+M_4)/2+k$\\
$\langle\stackrel{\text{ii}}{\bullet}\stackrel{\text{i}}{\bullet}\stackrel{1}{\circ}\stackrel{2}{\circ}\stackrel{4}{\circ}\cdot\stackrel{3}{\circ}\rangle$&$(M_1-k,M_2-k,M_3+k,M_4+k)$&$N+(-M_1-M_2+M_3+M_4)/2+k$\\
$\langle\stackrel{\text{ii}}{\bullet}\stackrel{\text{i}}{\bullet}\stackrel{1}{\circ}\stackrel{3}{\circ}\stackrel{4}{\circ}\cdot\stackrel{2}{\circ}\rangle$&$(M_1-k,M_2+k,M_3-k,M_4+k)$&$N+(-M_1+M_2-M_3+M_4)/2+k$\\
$\langle\stackrel{\text{ii}}{\bullet}\stackrel{\text{i}}{\bullet}\stackrel{2}{\circ}\stackrel{3}{\circ}\stackrel{4}{\circ}\cdot\stackrel{1}{\circ}\rangle$&$(M_1+k,M_2-k,M_3-k,M_4+k)$&$N+(M_1-M_2-M_3+M_4)/2+k$\\
$\langle\stackrel{\text{ii}}{\bullet}\cdot\stackrel{\text{i}}{\bullet}\stackrel{1}{\circ}\stackrel{2}{\circ}\stackrel{3}{\circ}\stackrel{4}{\circ}\rangle$&$(M_1,M_2,M_3,M_4+2k)$&$N+M_4+k$\\
$\langle\stackrel{\text{ii}}{\bullet}\stackrel{1}{\circ}\cdot\stackrel{\text{i}}{\bullet}\stackrel{2}{\circ}\stackrel{3}{\circ}\stackrel{4}{\circ}\rangle$&$(M_1-k,M_2+k,M_3+k,M_4+k)$&$N+(-M_1+M_2+M_3+M_4)/2+k$\\
$\langle\stackrel{\text{ii}}{\bullet}\stackrel{2}{\circ}\cdot\stackrel{\text{i}}{\bullet}\stackrel{1}{\circ}\stackrel{3}{\circ}\stackrel{4}{\circ}\rangle$&$(M_1+k,M_2-k,M_3+k,M_4+k)$&$N+(M_1-M_2+M_3+M_4)/2+k$\\
$\langle\stackrel{\text{ii}}{\bullet}\stackrel{3}{\circ}\cdot\stackrel{\text{i}}{\bullet}\stackrel{1}{\circ}\stackrel{2}{\circ}\stackrel{4}{\circ}\rangle$&$(M_1+k,M_2+k,M_3-k,M_4+k)$&$N+(M_1+M_2-M_3+M_4)/2+k$\\
$\langle\stackrel{\text{ii}}{\bullet}\stackrel{4}{\circ}\cdot\stackrel{\text{i}}{\bullet}\stackrel{1}{\circ}\stackrel{2}{\circ}\stackrel{3}{\circ}\rangle$&$(M_1-k,M_2-k,M_3-k,M_4+k)$&$N+(-M_1-M_2-M_3+M_4)/2+k$\\
$\langle\stackrel{\text{ii}}{\bullet}\stackrel{1}{\circ}\stackrel{2}{\circ}\cdot\stackrel{\text{i}}{\bullet}\stackrel{3}{\circ}\stackrel{4}{\circ}\rangle$&$(M_1,M_2,M_3+2k,M_4)$&$N+M_3+k$\\
$\langle\stackrel{\text{ii}}{\bullet}\stackrel{1}{\circ}\stackrel{3}{\circ}\cdot\stackrel{\text{i}}{\bullet}\stackrel{2}{\circ}\stackrel{4}{\circ}\rangle$&$(M_1,M_2+2k,M_3,M_4)$&$N+M_2+k$\\
$\langle\stackrel{\text{ii}}{\bullet}\stackrel{1}{\circ}\stackrel{4}{\circ}\cdot\stackrel{\text{i}}{\bullet}\stackrel{2}{\circ}\stackrel{3}{\circ}\rangle$&$(M_1-2k,M_2,M_3,M_4)$&$N-M_1+k$\\
$\langle\stackrel{\text{ii}}{\bullet}\stackrel{2}{\circ}\stackrel{3}{\circ}\cdot\stackrel{\text{i}}{\bullet}\stackrel{1}{\circ}\stackrel{4}{\circ}\rangle$&$(M_1+2k,M_2,M_3,M_4)$&$N+M_1+k$\\
$\langle\stackrel{\text{ii}}{\bullet}\stackrel{2}{\circ}\stackrel{4}{\circ}\cdot\stackrel{\text{i}}{\bullet}\stackrel{1}{\circ}\stackrel{3}{\circ}\rangle$&$(M_1,M_2-2k,M_3,M_4)$&$N-M_2+k$\\
$\langle\stackrel{\text{ii}}{\bullet}\stackrel{3}{\circ}\stackrel{4}{\circ}\cdot\stackrel{\text{i}}{\bullet}\stackrel{1}{\circ}\stackrel{2}{\circ}\rangle$&$(M_1,M_2,M_3-2k,M_4)$&$N-M_3+k$\\
$\langle\stackrel{\text{ii}}{\bullet}\stackrel{1}{\circ}\stackrel{2}{\circ}\stackrel{3}{\circ}\cdot\stackrel{\text{i}}{\bullet}\stackrel{4}{\circ}\rangle$&$(M_1+k,M_2+k,M_3+k,M_4-k)$&$N+(M_1+M_2+M_3-M_4)/2+k$\\
$\langle\stackrel{\text{ii}}{\bullet}\stackrel{1}{\circ}\stackrel{2}{\circ}\stackrel{4}{\circ}\cdot\stackrel{\text{i}}{\bullet}\stackrel{3}{\circ}\rangle$&$(M_1-k,M_2-k,M_3+k,M_4-k)$&$N+(-M_1-M_2+M_3-M_4)/2+k$\\
$\langle\stackrel{\text{ii}}{\bullet}\stackrel{1}{\circ}\stackrel{3}{\circ}\stackrel{4}{\circ}\cdot\stackrel{\text{i}}{\bullet}\stackrel{2}{\circ}\rangle$&$(M_1-k,M_2+k,M_3-k,M_4-k)$&$N+(-M_1+M_2-M_3-M_4)/2+k$\\
$\langle\stackrel{\text{ii}}{\bullet}\stackrel{2}{\circ}\stackrel{3}{\circ}\stackrel{4}{\circ}\cdot\stackrel{\text{i}}{\bullet}\stackrel{1}{\circ}\rangle$&$(M_1+k,M_2-k,M_3-k,M_4-k)$&$N+(M_1-M_2-M_3-M_4)/2+k$\\
$\langle\stackrel{\text{ii}}{\bullet}\stackrel{1}{\circ}\stackrel{2}{\circ}\stackrel{3}{\circ}\stackrel{4}{\circ}\cdot\stackrel{\text{i}}{\bullet}\rangle$&$(M_1,M_2,M_3,M_4-2k)$&$N-M_4+k$\\
$\langle\stackrel{1}{\circ}\cdot\stackrel{\text{ii}}{\bullet}\stackrel{\text{i}}{\bullet}\stackrel{2}{\circ}\stackrel{3}{\circ}\stackrel{4}{\circ}\rangle$&$(M_1-k,M_2+k,M_3+k,M_4-k)$&$N+(-M_1+M_2+M_3-M_4)/2+k$\\
$\langle\stackrel{2}{\circ}\cdot\stackrel{\text{ii}}{\bullet}\stackrel{\text{i}}{\bullet}\stackrel{1}{\circ}\stackrel{3}{\circ}\stackrel{4}{\circ}\rangle$&$(M_1+k,M_2-k,M_3+k,M_4-k)$&$N+(M_1-M_2+M_3-M_4)/2+k$\\
$\langle\stackrel{3}{\circ}\cdot\stackrel{\text{ii}}{\bullet}\stackrel{\text{i}}{\bullet}\stackrel{1}{\circ}\stackrel{2}{\circ}\stackrel{4}{\circ}\rangle$&$(M_1+k,M_2+k,M_3-k,M_4-k)$&$N+(M_1+M_2-M_3-M_4)/2+k$\\
$\langle\stackrel{4}{\circ}\cdot\stackrel{\text{ii}}{\bullet}\stackrel{\text{i}}{\bullet}\stackrel{1}{\circ}\stackrel{2}{\circ}\stackrel{3}{\circ}\rangle$&$(M_1-k,M_2-k,M_3-k,M_4-k)$&$N+(-M_1-M_2-M_3-M_4)/2+k$\\
\end{tabular}
\end{center}
\caption{Changes of the relative ranks $(M_1,M_2,M_3,M_4)$ and the overall rank $N$, for the brane configurations of the $F_4$ Weyl group.
Here we change references into that denoted by $\cdot$ in applying duality cascades.}
\label{changerefF4}
\end{table}

Again, we would like to identify the processes of duality cascades as translations in the parameter space of the relative ranks $\{(M_1,M_2,M_3,M_4)\}$ and the overall rank $N$ for the brane configurations \eqref{F4braneconfig}
\begin{align}
&\Big\langle N\stackrel{\text{ii}}{\bullet}N+M_4+k\stackrel{1}{\circ}
N+\frac{1}{2}(-M_1+M_2+M_3+M_4)+k\stackrel{2}{\circ}
N+M_3+k\stackrel{\text{i}}{\bullet}\nonumber\\
&N+M_3+M_4+2k\stackrel{3}{\circ}
N+\frac{1}{2}(M_1+M_2+M_3+M_4)+k\stackrel{4}{\circ}
\Big\rangle.
\label{f4M}
\end{align}
This, however, has to be done with caution due to the constraint \eqref{balance}.
If the order of the two NS5-branes is exchanged in changing references, we also need to relabel them simultaneously.
For example, if we change references into $\langle\stackrel{\text{ii}}{\bullet}\cdot\stackrel{\text{i}}{\bullet}\stackrel{1}{\circ}\stackrel{2}{\circ}\stackrel{3}{\circ}\stackrel{4}{\circ}\rangle$, we relabel the two NS5-branes $\stackrel{\text{ii}}{\bullet}$ and $\stackrel{\text{i}}{\bullet}$ as
\begin{align}
&\Big\langle N+M_4+k\stackrel{1}{\circ}
N+\frac{1}{2}(-M_1+M_2+M_3+M_4)+k\stackrel{2}{\circ}
N+M_3+k\stackrel{\text{ii}}{\bullet}\nonumber\\
&N+M_3+M_4+2k\stackrel{3}{\circ}
N+\frac{1}{2}(M_1+M_2+M_3+M_4)+k\stackrel{4}{\circ}
N\stackrel{\text{i}}{\bullet}\Big\rangle.
\end{align}
By returning to the original order of 5-branes, we find the variable change
$(M_1,M_2,M_3,M_4)\to(M_1,M_2,M_3,M_4+2k)$ and $N\to N+M_4+k$.
Similarly, we have performed the analysis for various changes of references.
We list the results in table \ref{changerefF4}.

As in the $B_3$ case, the change of references into $\langle\stackrel{4}{\circ}\cdot\stackrel{\text{ii}}{\bullet}\stackrel{\text{i}}{\bullet}\stackrel{1}{\circ}\stackrel{2}{\circ}\stackrel{3}{\circ}\rangle$ is realized by the translation
\begin{align}
t_0=s_0s_4s_3s_{25}s_{16}s_3s_{25}s_3s_4s_3s_{25}s_3s_{16}s_{25}s_3s_4,
\end{align}
in the affine Weyl group.
Here the affine Weyl reflection $s_0$ for the relative ranks is given by
\begin{align}
s_0&:(M_1,M_2,M_3,M_4)\nonumber\\
&\quad\mapsto\bigg(\frac{M_1-M_2-M_3-M_4}{2}+k,\frac{-M_1+M_2-M_3-M_4}{2}+k,\frac{-M_1-M_2+M_3-M_4}{2}+k,
\nonumber\\
&\qquad\qquad\qquad\frac{-M_1-M_2-M_3+M_4}{2}+k\bigg),
\end{align}
where all the points in \eqref{pointsF4} are kept fixed, while that for the overall rank is
\begin{align}
s_0:N\mapsto N-\frac{M_1+M_2+M_3+M_4}{2}+k,
\end{align}
which is equivalent to \eqref{Nreflection}.
After equipped with the affine Weyl reflection $s_0$, translations in other directions can also be generated in the affine Weyl group.

In summary we have found that, as in the $B_3$ case, the duality cascades are realized as translations in the parameter space for brane configurations and these translations are generated by the affine $\widehat F_4$ Weyl group.
Again, the affine Weyl group guarantees that the fundamental domain forms a parallelotope, which ensures the physical properties we expect for the duality cascades.

\section{Deformations with FI parameters}\label{FI}

In the previous section, we have studied the fundamental domain of duality cascades and the processes of them.
Among others, we have found that the fundamental domain in the Weyl chamber is nothing but the affine Weyl chamber.
These studies are, however, confined to the $B_3$ subspace of the full $D_5$ space or the $F_4$ subspace of the full $E_7$ space.
To return to the full space, we need to introduce FI parameters, which are realized in brane configurations as the position shifts in the perpendicular directions.
It is known that, for the cases without rank differences, the FI parameters are realized by shifting the canonical operators as
\begin{align}
\widehat H^{-1}=\prod_{a=1}^R\bigg(2\cosh\frac{\widehat r-2\pi iZ_a}{2}\bigg)^{-1},
\label{Hinverse}
\end{align}
where $\widehat r$ denotes $\widehat p$ for NS5-branes and $\widehat q$ for $(1,k)$5-branes and $R$ is the total number of the 5-branes.
As previously without deformations of FI parameters \cite{KM,FMN}, we can extrapolate to the cases with rank differences by considering various orders of 5-branes, so that the full parameters of the curves are realized.

In section \ref{FDDC}, we have considered the fundamental domain of duality cascades by requiring the reference rank to be lower than others in brane configurations obtained from the HW transitions.
After introducing the FI parameters, however, the HW transitions are not enough to generate the whole system of inequalities for the fundamental domain.
Fortunately, since the corresponding curve enjoys symmetries of the Weyl groups, we can utilize the Weyl groups to generate inequalities for the $D_5$ and $E_7$ quantum curves.
After that we interpret symmetries of the Weyl group as brane transitions.

\subsection{$D_5$}\label{sectiond5}

Now let us introduce the FI parameters $Z_i$ to the brane configurations \eqref{M123},
\begin{align}
&\langle N_1\overunderset{}{Z_2}{\bullet}N_2\overunderset{}{Z_1}{\bullet}N_3\overunderset{}{Z_3}{\circ}N_4\overunderset{}{Z_4}{\circ}\rangle
\nonumber\\
&=\langle M_2+M_3\overunderset{}{Z_2}{\bullet}M_1+2M_3+k\overunderset{}{Z_1}{\bullet}2M_1+M_2+M_3+2k\overunderset{}{Z_3}{\circ}M_1+k\overunderset{}{Z_4}{\circ}\rangle,
\label{braneD5}
\end{align}
where we omit the overall rank $N_0$ and the labels of $\bullet$ and $\circ$ as well since we cannot trace the labels any more in the following studies.

To study the fundamental domain of duality cascades later, let us relate the variables between brane configurations and quantum curves.
With the insertions of the FI parameters, the spectral operator is changed and we can cover the whole parameter space of the $D_5$ quantum curve.
As in \eqref{Hinverse}, the spectral operator is given by
\begin{align}
\widehat H=\bigg(2\cosh\frac{\widehat q-2\pi iZ_4}{2}\bigg)
\bigg(2\cosh\frac{\widehat q-2\pi iZ_3}{2}\bigg)
\bigg(2\cosh\frac{\widehat p-2\pi iZ_1}{2}\bigg)
\bigg(2\cosh\frac{\widehat p-2\pi iZ_2}{2}\bigg).
\end{align}
Since the commutation relations
\begin{align}
&\widehat P^{\frac{n}{2}}
(e^{-\pi iZ}\widehat Q^{\frac{1}{2}}
+e^{\pi iZ}\widehat Q^{-\frac{1}{2}})
=(q^{-\frac{n}{4}}e^{-\pi iZ}\widehat Q^{\frac{1}{2}}
+q^{\frac{n}{4}}e^{\pi iZ}\widehat Q^{-\frac{1}{2}})
\widehat P^{\frac{n}{2}},\nonumber\\
&(e^{-\pi iZ}\widehat P^{\frac{1}{2}}
+e^{\pi iZ}\widehat P^{-\frac{1}{2}})
\widehat Q^{\frac{n}{2}}
=\widehat Q^{\frac{n}{2}}
(q^{-\frac{n}{4}}e^{-\pi iZ}\widehat P^{\frac{1}{2}}
+q^{\frac{n}{4}}e^{\pi iZ}\widehat P^{-\frac{1}{2}}),
\end{align}
imply that the appearance of the factor $q$ is the same as in the case without FI parameters \eqref{commutation}, the asymptotic values \eqref{D5asympt} (with the conditions \eqref{d5constraint} and \eqref{hem} chosen) are simply shifted by $z_i=e^{2\pi iZ_i}$,
\begin{align}
\bigg(\frac{1}{e_1},\frac{1}{e_2};e_3,e_4;\frac{e_5}{h_2},\frac{e_6}{h_2};\frac{h_1}{e_7},\frac{h_1}{e_8}\bigg)
=\bigg(\frac{m_2z_1}{m_3},1;m_2m_3z_3,1;\frac{z_1}{m_1},\frac{m_2}{m_1m_3};\frac{z_3}{m_1},\frac{m_2m_3}{m_1}\bigg).
\end{align}
Here we have chosen $Z_2=Z_4=0$ by fixing the origin for the perpendicular directions or
\begin{align}
z_2=z_4=1,
\label{sdetcond}
\end{align}
corresponding to $e_2=e_4=1$ \eqref{d5constraint}.
Then, we can match the parameters of the brane configurations with the parameters of the $D_5$ curve as
\begin{align}
(h_1,h_2,e_1,e_3,e_5)
=\bigg(\frac{m_2m_3}{m_1},\frac{m_1m_3}{m_2},\frac{m_3}{m_2z_1},m_2m_3z_3,\frac{m_3z_1}{m_2}\bigg).
\label{hemz}
\end{align}
Using the variables \eqref{hemz}, the transformations of the $D_5$ Weyl group are given as \eqref{D5transf}.

After introducing FI parameters and understanding their transformations, we can study the fundamental domain of duality cascades.
Previously, we require the condition that the reference rank continues to be the lowest in applying the HW transitions arbitrarily.
Since the whole system of inequalities enjoys symmetries of the $B_3$ Weyl group, we can alternatively obtain those inequalities by applying the transformations \eqref{sM} to the original ones
\begin{align}
M_2+M_3\le M_1+2M_3+k,\quad
M_2+M_3\le 2M_1+M_2+M_3+2k,\quad
M_2+M_3\le M_1+k.
\label{M123ineq}
\end{align}
Hence, for the deformation with the FI parameters, instead of applying the HW transitions, we can apply the Weyl group to the inequalities to find out the fundamental domain similarly.

Namely, we first rewrite the transformations \eqref{D5transf} in terms of $M_i$ and $Z_i$ with $m_i=e^{2\pi iM_i}$ and $z_i=e^{2\pi iZ_i}$ (by changing multiplications into additions) and then apply the transformations to \eqref{M123ineq} to generate a system of inequalities in five dimensions.
The resulting condition is
\begin{align}
&\pm M_1\le k,\quad
\pm M_2\le k/2,\quad
\pm M_3\le k/2,\quad
\pm M_1\pm M_2\pm M_3\le k,\nonumber\\
&\pm M_2\pm M_3\le k,\quad
\pm Z_1\le k,\quad
\pm Z_3\le k,\quad
\pm Z_1\pm Z_3\le k,\nonumber\\
&\pm M_1\pm Z_1\le k,\quad
\pm M_1\pm Z_3\le k,\quad
\pm M_2\pm M_3\pm Z_1\le k,\quad
\pm M_2\pm M_3\pm Z_3\le k,
\label{domainD5}
\end{align}
where the inequalities on the first line are those describing the rhombic dodecahedron in \eqref{ineqB3trivial}.
As in \eqref{ineqB3trivial}, some of them are trivially guaranteed from the others.
For example, inequalities $\pm M_2\pm M_3\le k$ are obtained from $\pm M_2\le k/2$ and $\pm M_3\le k/2$, while inequalities $\pm Z_1\le k$ and $\pm Z_3\le k$ are obtained from $\pm Z_1\pm Z_3\le k$.
By removing them, we are left with 40 inequalities
\begin{align}
&\pm M_2\le k/2,\quad\pm M_3\le k/2,\quad\pm M_1\pm M_2\pm M_3\le k,\quad
\pm Z_1\pm Z_3\le k,\nonumber\\
&\pm M_1\pm Z_1\le k,\quad
\pm M_1\pm Z_3\le k,\quad
\pm M_2\pm M_3\pm Z_1\le k,\quad
\pm M_2\pm M_3\pm Z_3\le k.
\label{ineqD5}
\end{align}

Since the inequalities \eqref{ineqD5} are generated by the $D_5$ Weyl group, it is not difficult to imagine that we may not need the whole three inequalities in \eqref{M123ineq} to generate \eqref{ineqD5}.
Indeed, the 40 inequalities in \eqref{ineqD5} can be generated only by the first inequality or the third one in \eqref{M123ineq}.
(Since the second inequality is trivially guaranteed from the others in \eqref{ineqB3trivial}, this only generates trivial ones.)
It is also interesting to note that the same system of inequalities can alternatively be generated from the condition for FI parameters, one of $\pm Z_1\pm Z_3\le k$, instead of the original condition for ranks.

Since the system is generated from the $D_5$ Weyl group, it trivially enjoys symmetries of Weyl reflections \eqref{reflection} in the space $\{\lambda=(M_1,M_2,M_3,Z_1,Z_3)\}$ with the metric for the bilinear form and the simple roots $\alpha_i$ given by $g=\diag(1,2,2,1,1)$ and \eqref{pointD5}.
We can then study the fundamental domain with the same method by restricting ourselves to the $D_5$ Weyl chamber
\begin{align}
&(\alpha_1,\lambda)=M_2+M_3-Z_3\ge 0,\quad
(\alpha_2,\lambda)=M_2+M_3+Z_3\ge 0,\nonumber\\
&(\alpha_3,\lambda)=-M_1-M_2-M_3\ge 0,\quad
(\alpha_4,\lambda)=M_1+M_2-M_3\ge 0,\nonumber\\
&(\alpha_5,\lambda)=-M_2+M_3-Z_1\ge 0.
\end{align}
As previously, we find intersections with the fundamental domain \eqref{domainD5} to be
\begin{align}
k\omega_1/2,\quad k\omega_2/2,\quad
k\omega_3,\quad k\omega_4,\quad
k\omega_5/2,
\label{d5intersect}
\end{align}
(with $\omega_i$ being the fundamental weights \eqref{D5fw}), where all of them lie on the hyperplane $M_2-M_3-Z_1=k$.
Since the inequality
\begin{align}
M_2-M_3-Z_1\le k,
\label{boundaryD5}
\end{align}
already appears in the condition \eqref{domainD5}, the fundamental domain in the $D_5$ Weyl chamber is the convex hull of the five points in \eqref{d5intersect} and the origin.

Interestingly, the inequality \eqref{boundaryD5} can be expressed as \eqref{affineineq} as in the $B_3$ case without the deformation by FI parameters, if we introduce the highest root as
\begin{align}
\theta=(0,1/2,-1/2,-1,0)=\alpha_1+\alpha_2+2\alpha_3+2\alpha_4+\alpha_5.
\end{align}
Note that, although we have imposed the $D_5$ Weyl group by hand, we have not imposed the affine $\widehat D_5$ Weyl group.
Nevertheless, in studying the fundamental domain of duality cascades with the deformation by the FI parameters, the affine $\widehat D_5$ Weyl chamber appears naturally.
Here the $\widehat D_5$ Dynkin diagram can be found in figure \ref{d5dynkin} in appendix \ref{appendixD5}.
Later we shall try to present the brane interpretation of imposing the $D_5$ Weyl group for the fundamental domain.
Before discussing the interpretation, let us repeat the same analysis for the $E_7$ case.

\subsection{$E_7$}\label{sectione7}

As in the $D_5$ case, let us consider FI parameters for the $F_4$ case and study the fundamental domain similarly by generating inequalities with the Weyl group.
We introduce the FI parameters $Z_i$ to the brane configurations \eqref{f4M}
\begin{align}
&\Big\langle 0\overunderset{}{0}{\bullet}M_4+k\overunderset{}{Z_1}{\circ}\frac{1}{2}(-M_1+M_2+M_3+M_4)+k\overunderset{}{Z_2}{\circ}\nonumber\\
&M_3+k\overunderset{}{0}{\bullet}M_3+M_4+2k\overunderset{}{Z_3}{\circ}\frac{1}{2}(M_1+M_2+M_3+M_4)+k\overunderset{}{0}{\circ}\Big\rangle,
\label{E7braneconfig}
\end{align}
with the overall rank $N$ and the labels of 5-branes abbreviated.
As in the $F_4$ case, by requiring the reference rank to be the lowest one, we find four inequalities
\begin{align}
&M_4+k\ge 0,\quad
\frac{1}{2}(-M_1+M_2+M_3+M_4)+k\ge 0,\nonumber\\
&M_3+k\ge 0,\quad
\frac{1}{2}(M_1+M_2+M_3+M_4)+k\ge 0.
\label{fourineq}
\end{align}
Note that $M_3+M_4+2k\ge 0$ is trivially guaranteed from $M_4+k\ge 0$ and $M_3+k\ge 0$.
In the following we apply the $E_7$ Weyl group to find out the whole system of inequalities.

For this purpose, let us clarify the change of parameters first.
In the original variables $(f_1,f_2,f_3,g_1,h_1,h_2,h_3)$ with the condition \eqref{fghcondition} chosen, the asymptotic values \eqref{E7asympt} are
\begin{align}
(g_1\text{[double]};f_1,f_2,f_3,1;1\text{[double]};(g_1h_1)^{-1},(g_1h_2)^{-1},(g_1h_3)^{-1},f_1f_2f_3g_1h_1h_2h_3).
\label{e7asymptotic}
\end{align}
After restricting to the $F_4$ subspace \eqref{F4subspace}, applying the variable changes \eqref{M1234},
\begin{align}
f_1=(m_2m_3)^{-\frac{1}{2}},\quad
f_2=(m_1m_3)^{-\frac{1}{2}},\quad
f_3=(m_1m_2)^{-\frac{1}{2}},\quad
g_1=m_4,
\end{align}
and introducing the FI parameters by shifting the asymptotic values by $z_i$,
\begin{align}
&(m_4\text{[double]};z_1(m_2m_3)^{-\frac{1}{2}},z_2(m_1m_3)^{-\frac{1}{2}},z_3(m_1m_2)^{-\frac{1}{2}},1;\nonumber\\
&1\text{[double]};z_1(m_1m_4)^{-\frac{1}{2}},z_2(m_2m_4)^{-\frac{1}{2}},z_3(m_3m_4)^{-\frac{1}{2}},(m_1m_2m_3m_4)^{-\frac{1}{2}}),
\label{asymptotic1234}
\end{align}
we find that the variables are identified as
\begin{align}
(f_1,f_2,f_3,g_1,h_1,h_2,h_3)
=\bigg(\frac{z_1}{\sqrt{m_2m_3}},\frac{z_2}{\sqrt{m_1m_3}},\frac{z_3}{\sqrt{m_1m_2}},m_4,
\frac{1}{z_1}\sqrt{\frac{m_1}{m_4}},\frac{1}{z_2}\sqrt{\frac{m_2}{m_4}},\frac{1}{z_3}\sqrt{\frac{m_3}{m_4}}\bigg),
\label{m1234z123}
\end{align}
by comparing \eqref{asymptotic1234} with the original parametrization \eqref{e7asymptotic}.

We can then apply the transformations of the $E_7$ Weyl group in appendix \ref{appendixE7} to the inequalities \eqref{fourineq} and find the condition
\begin{align}
&\pm M_i\le k,\quad
\pm M_4\le k,\quad
\pm M_1\pm M_2\pm M_3\pm M_4\le 2k,\nonumber\\
&\pm M_i\pm M_j\pm 2Z_k\le 2k,\quad
\pm M_i\pm M_4\pm 2Z_i\le 2k,\nonumber\\
&\pm M_i\pm M_j\pm 2(Z_i-Z_j)\le 2k,\quad
\pm M_i\pm M_4\pm 2(Z_j-Z_k)\le 2k,\nonumber\\
&\pm Z_1\pm Z_2\pm Z_3\le k,
\label{ineqE7}
\end{align}
for the fundamental domain.
Here the indices $(i,j,k)$ stand for 1,2,3 which are different in the same equations.
Also, all combinations of multiple signs are valid except the last one;
the combinations $\pm(Z_1+Z_2+Z_3)\le k$ with all the same sign are absent.
The condition consists of $3\times 2+2+2^4+{}_3C_2\times 2^3+{}_3C_1\times 2^3+{}_3C_2\times 2^3+{}_3C_1\times 2^3+(2^3-2)=126$ inequalities in total.
Note that the 24 inequalities independent of the FI parameters $Z_i$ on the first line are those describing the 24-cell in \eqref{24cellineq}.

Since we apply the $E_7$ Weyl group to generate the inequalities, apparently the fundamental domain enjoys symmetries of the $E_7$ Weyl group.
Using the Weyl group in appendix \ref{appendixE7}, we can reduce the fundamental domain to the Weyl chamber $(\alpha_i,\lambda)\ge 0$ with the simple roots $\alpha_i$.
Then, we can study intersections with the fundamental domain, which are
\begin{align}
k\omega_1/2,\quad
k\omega_2/3,\quad
k\omega_3/4,\quad
k\omega_4/2,\quad
k\omega_5/3,\quad
k\omega_6/2,\quad
k\omega_7,
\label{pointE7}
\end{align}
(with $\omega_i$ being the fundamental weights) staying on the hyperplane $M_1+M_2-2Z_3=2k$.
Since the inequality
\begin{align}
M_1+M_2-2Z_3\le 2k,
\label{boundaryE7}
\end{align}
appears in \eqref{ineqE7}, the fundamental domain is the convex hull of the points in \eqref{pointE7} and the origin.
Again, the inequality \eqref{boundaryE7} can be rewritten as \eqref{affineineq}, if we introduce the highest root
\begin{align}
\theta=2\alpha_1+3\alpha_2+4\alpha_3+2\alpha_4+3\alpha_5+2\alpha_6+\alpha_7,
\end{align}
which may not be surprising by now.
The $\widehat E_7$ Dynkin diagram can be found in figure \ref{e7dynkin} in appendix \ref{appendixE7}.

In summary, for both the $D_5$ case and the $E_7$ case, we introduce the FI parameters to detect the whole parameter space of the curves.
We require the same condition that the reference rank is the lowest to study the fundamental domain of duality cascades.
Previously without deformations, we have applied the HW transitions to generate inequalities for the fundamental domain.
Here, although we do not have the explicit rules for brane transitions, from symmetries of the Weyl group for the quantum curve, we generate the whole system of inequalities and find that the fundamental domain in the Weyl chamber is again the affine Weyl chamber.
Hence, it is natural to expect that the brane configurations continue to enjoy symmetries of the affine Weyl group even after we introduce the deformation with the FI parameters.

\subsection{Half Hanany-Witten brane transitions}\label{sectionhalf}

In the previous subsections, we have introduced FI parameters to discuss quantum curves with full parameters.
Using the full Weyl group without the ${\mathbb Z}_2$ folding, we have generated inequalities to define the fundamental domain, though the transformations cannot be interpreted as the HW transitions.
Here we shall discuss their interpretation in brane configurations.

First let us start with the $D_5$ quantum curve.
Without deformations of FI parameters, the combined transformations $s_{12}=s_1s_2$ and $s_{50}=s_5s_0$ were interpreted as the trivial HW transitions exchanging 5-branes of the same type \cite{KM}.
After introducing the FI parameters, using \eqref{D5transf} they turn out to be
\begin{align}
s_1s_2&:(M_1,M_2,M_3,Z_1,Z_3)\mapsto(M_1,-M_3,-M_2,Z_1,-Z_3),\nonumber\\
s_5s_0&:(M_1,M_2,M_3,Z_1,Z_3)\mapsto\big(M_1,M_3,M_2,-Z_1,Z_3).
\label{s1250MZ}
\end{align}
It is natural to continue to interpret them as the HW transitions for the following reason.
Since we naturally expect that the HW transitions do not affect the FI parameters in the perpendicular directions, the exchanges of 5-branes of each type simply induce the transformations $Z_1\leftrightarrow Z_2$ and $Z_3\leftrightarrow Z_4$ in \eqref{braneD5}, which reduce to $Z_1\leftrightarrow -Z_1$ and $Z_3\leftrightarrow -Z_3$ respectively after fixing the condition \eqref{sdetcond}.
Then, this matches with the transformations in \eqref{s1250MZ} completely.
Here, besides the HW transitions, we also have ``halves'' of them ($s_1$, $s_2$, $s_5$ and $s_0$) in \eqref{D5transf}.
By rewriting the transformations \eqref{D5transf} in terms of the brane configurations \eqref{braneD5}, we find that these transformations map \eqref{braneD5} into
\begin{align}
s_1&:\mapsto
\langle M_2+M_3\overunderset{}{0}{\bullet}M_1+2M_3+k\overunderset{}{Z_1}{\bullet}2M_1+M_2+M_3+2k\overunderset{}{M_2+M_3}{\circ}M_1+M_2+M_3-Z_3+k\overunderset{}{0}{\circ}\rangle,
\nonumber\\
s_2&:\mapsto
\langle M_2+M_3\overunderset{}{0}{\bullet}M_1+2M_3+k\overunderset{}{Z_1}{\bullet}2M_1+M_2+M_3+2k\overunderset{}{-M_2-M_3}{\circ}M_1+M_2+M_3+Z_3+k\overunderset{}{0}{\circ}\rangle,
\nonumber\\
s_5&:\mapsto
\langle M_2+M_3\overunderset{}{0}{\bullet}M_1+M_2+M_3+Z_1+k\overunderset{}{-M_2+M_3}{\bullet}2M_1+M_2+M_3+2k\overunderset{}{Z_3}{\circ}M_1+k\overunderset{}{0}{\circ}\rangle,
\nonumber\\
s_0&:\mapsto
\langle M_2+M_3\overunderset{}{0}{\bullet}M_1+M_2+M_3-Z_1+k\overunderset{}{M_2-M_3}{\bullet}2M_1+M_2+M_3+2k\overunderset{}{Z_3}{\circ}M_1+k\overunderset{}{0}{\circ}\rangle.
\label{sbraneconfig}
\end{align}

Also, for the $E_7$ case, the combined transformations are given by
\begin{align}
s_0s_7&:(M_1,M_2,M_3,M_4,Z_1,Z_2,Z_3)
\mapsto(-M_2,-M_1,M_3,M_4,Z_1-Z_3,Z_2-Z_3,-Z_3),\nonumber\\
s_1s_6&:(M_1,M_2,M_3,M_4,Z_1,Z_2,Z_3)
\mapsto(M_1,M_3,M_2,M_4,Z_1,Z_3,Z_2),\nonumber\\
s_2s_5&:(M_1,M_2,M_3,M_4,Z_1,Z_2,Z_3)
\mapsto(M_2,M_1,M_3,M_4,Z_2,Z_1,Z_3).
\label{sMZe7}
\end{align}
Since the FI parameters $Z_1$, $Z_2$ and $Z_3$ transform trivially in exchanging the 5-branes, we can associate \eqref{sMZe7} as the HW transitions exchanging 5-branes of the same type.
Then, again, the separated ones $s_0$, $s_7$, $s_1$, $s_6$, $s_2$ and $s_5$ can be interpreted as ``halves'' of the HW transitions.
As \eqref{sbraneconfig} in the $D_5$ case, we can spell out the transformations explicitly in brane configurations.

By comparing the halves of the HW transitions for the $D_5$ curve and the $E_7$ curve, we find that the brane transitions can be summarized by the rule
\begin{align}
s_\pm:\bigg\langle\cdots K\overunderset{}{X}{\left[\begin{matrix}\bullet\\\circ\end{matrix}\right]}L\overunderset{}{Y}{\left[\begin{matrix}\bullet\\\circ\end{matrix}\right]}M\cdots\bigg\rangle\mapsto\bigg\langle
\cdots K\overunderset{}{\frac{X+Y}{2}\mp\frac{K-2L+M}{4}}{\left[\begin{matrix}\bullet\\\circ\end{matrix}\right]}\frac{K+M}{2}\pm(X-Y)\overunderset{}{\frac{X+Y}{2}\pm\frac{K-2L+M}{4}}{\left[\begin{matrix}\bullet\\\circ\end{matrix}\right]}M\cdots\bigg\rangle.
\label{transition}
\end{align}
Here note that the brane transitions $s_+$ and $s_-$ are commutative and combine into the HW transition $s_+s_-=s_-s_+$.
Also, interestingly, the two transitions $s_+$ and $s_-$ are simply related by reversing left and right for the brane configurations.

After rewriting the transformations of the Weyl groups into brane transitions, it can be understood that duality cascades are generated by these brane transitions.
Namely, compared with the cases without deformations in section \ref{FDDC} where duality cascades are generated by the HW transitions, here duality cascades are generated even for the transitions \eqref{transition} in addition to the original HW transitions.
We still change references even when we encounter lower ranks in applications of the transitions \eqref{transition}.
 
Of course, at present we do not have any deep understanding for the transitions \eqref{transition} mixing the relative ranks with the FI parameters.
However, we stress that the transitions are not imposed by hand.
Instead, they originate naturally from the Weyl group.
Before closing this section, let us try to give another interpretation for the FI parameters as super determinant operators.

\subsection{Super determinant operators}

In this section, we have studied the fundamental domain of duality cascades corresponding to quantum curves with full parameters by introducing FI parameters and identified the fundamental domains by imposing the inequalities generated by the Weyl groups.
In the previous subsection we have clarified the requirement of the Weyl groups by translating the transformations into brane transitions.
Note that the FI parameters are closely related to mass deformations and appear in various contexts such as phase transitions \cite{AZ,AR,NST,HNST}.
Before closing this section, we shortly comment on another interpretation for the FI parameters.
The interpretation may already be known partially.
Our comments aim at future possible connections to integrabilities.

As is clear from its Fermi gas formalism, for the ${\cal N}=4$ supersymmetric Chern-Simons theories with gauge group $\prod_{a=1}^R\text{U}(N)_{k_a}$ (and without rank differences), the spectral operator \eqref{Hinverse} with the FI parameters can be rewritten as
\begin{align}
\widehat H^{-1}
=e^{-\frac{i}{2\hbar}\widehat p^2}\prod_{a=1}^R\bigg(
e^{-\frac{i}{2\hbar}s_a\widehat q^2}e^{-\frac{2\pi}{\hbar}Z_a\widehat q}
\frac{1}{2\cosh\frac{\widehat p}{2}}
e^{\frac{2\pi}{\hbar}Z_a\widehat q}e^{\frac{i}{2\hbar}s_a\widehat q^2}
\bigg)e^{\frac{i}{2\hbar}\widehat p^2},
\end{align}
with suitable applications of similarity transformations
\begin{align}
e^{-\frac{i}{2\hbar}\widehat q^2}f(\widehat q,\widehat p)e^{\frac{i}{2\hbar}\widehat q^2}
=f(\widehat q,\widehat p+\widehat q),\quad
e^{-\frac{i}{2\hbar}\widehat p^2}f(\widehat q,\widehat p)e^{\frac{i}{2\hbar}\widehat p^2}
=f(\widehat q-\widehat p,\widehat p).
\end{align}
Here the Chern-Simons levels are given by \cite{IK4}
\begin{align}
k_a=-k(s_a-s_{a-1}),
\label{ks}
\end{align}
where the binary $s_a$ denotes the type of the $a$-th 5-brane, with $s_a=0$ being the NS5-brane and $s_a=1$ being the $(1,k)$5-brane respectively under the identification $s_0=s_R$.
Similarly, we can combine the FI parameters into
\begin{align}
\zeta_a=Z_a-Z_{a-1}.
\end{align}
Then, it seems that we are considering the insertion of the operators
\begin{align}
\Delta(e^{\lambda_{a-1}},e^{\lambda_{a}})
=e^{-\left(\sum_{m=1}^{N_{a-1}}\lambda_{a-1,m}-\sum_{n=1}^{N_{a}}\lambda_{a,n}\right)},
\label{sdet}
\end{align}
into the generalized ABJM matrix models with multiplicity $\zeta_a$.
Especially for the ABJM case, the operator reduces to
\begin{align}
\Delta(e^\mu,e^\nu)=e^{-\left(\sum_{m=1}^{N_1}\mu_m-\sum_{n=1}^{N_2}\nu_n\right)}.
\label{ABJMsdet}
\end{align}

Returning to the original ABJM theory, it is known that the ABJM matrix model formally enjoys a hidden structure of supergroups $\text{U}(N_1|N_2)$ \cite{DT,MPts,KWY}.
Concretely, let us consider one-point correlation functions of the half-BPS Wilson loop operators (labeled by the Young diagram $\lambda$)
\begin{align}
\langle s_\lambda\rangle^\text{ABJM}_k(N_1|N_2)&=\frac{1}{N_1!N_2!}\int\frac{d^{N_1}\mu}{(2\pi)^{N_1}}\frac{d^{N_2}\nu}{(2\pi)^{N_2}}
e^{\frac{ik}{4\pi}(\sum_{m=1}^{N_1}\mu_m^2-\sum_{n=1}^{N_2}\nu_n^2)}\nonumber\\
&\qquad\times\frac{\prod_{m<m'}^{N_1}(2\sinh\frac{\mu_m-\mu_{m'}}{2})^2\prod_{n<n'}^{N_2}(2\sinh\frac{\nu_n-\nu_{n'}}{2})^2}
{\prod_{m=1}^{N_1}\prod_{n=1}^{N_2}(2\cosh\frac{\mu_m-\nu_n}{2})^2}\times s_\lambda(e^\mu,e^\nu).
\end{align}
Then, the integration measure can be regarded as the Chern-Simons deformation $x_m\to e^{\pm\mu_m}$, $y_n\to e^{\pm\nu_n}$ from that of the supergroup $\text{U}(N_1|N_2)$, $\prod_{m<m'}^{N_1}(x_m-x_{m'})\prod_{n<n'}^{N_2}(y_n-y_{n'})/\prod_{m=1}^{N_1}\prod_{n=1}^{N_2}(x_m+y_n)$, and the exponent as the supertrace. 
The half-BPS Wilson loop operators preserve the structure of the supergroup \cite{DT} and were found to reduce to the super Schur polynomials $s_\lambda(e^\mu,e^\nu)$.
In \cite{KM2pt} two-point correlation functions of the super Schur polynomials, $\langle s_\lambda s_{\lambda'}\rangle^\text{ABJM}_k(N_1|N_2)$ were proposed and studied carefully, which are expected to be obtained from two-point functions of the half-BPS Wilson loops in the ABJM theory.

The super Schur polynomials appear naturally because of the hidden structure of the supergroup $\text{U}(N_1|N_2)$.
With this structure, it is natural to ask why we have not considered the super determinant operator \eqref{ABJMsdet} associated to each 5-brane.
Similarly, we can generalize the ABJM theory to various ${\cal N}=4$ supersymmetric Chern-Simons theories of circular quivers.
The circular quiver inherits the structure of the supergroup $\text{U}(N_1|N_2)$ locally and allows the insertions of various super determinant operators \eqref{sdet} for each 5-brane.
Thus we can interpret FI parameters as super determinant operators.

It is interesting to note that the ABJM matrix model with the half-BPS Wilson loops satisfies various identifies such as the Giambelli identity and the Jacobi-Trudi identity which indicate the integrability structure \cite{HHMO,MM,G,JT,2DTL}.
This is why we attempt to relate the deformations of FI parameters to super determinant operators, which may be clarified from integrabilities.

In the ${\cal N}=4$ super Yang-Mills theory it was known that the determinant plays an important role in relating trace operators to string states \cite{BBNS,CJR}.
Similarly, here we expect that the super determinant operators originate from half-BPS composite or solitonic operators in the ABJM theory and are crucial to clarify the M2-branes.

\section{Discussions and future directions}

Let us first summarize the content of this paper.
In this paper we reconsider the HW brane transitions in a circle $S^1$.
As previously in \cite{MNN,KM}, it was proposed that in defining the grand canonical partition functions and comparing with brane configurations, we need to fix the reference frame which the 5-branes do not move across and sum over all non-vanishing partition functions with non-negative ranks.
Also, in the so-called closed formalism proposed in \cite{H1,KMZ,MS2,MN5,closed}, it is natural to regard the lowest rank as full-fledged M2-branes and consider fractional M2-branes as fluctuations to be integrated out.
We point out that these manipulations and viewpoints of the grand canonical partition functions are consistent with the proposed duality cascades in \cite{KS,HK}.
Namely, we propose the explicit process of duality cascades by continuing to assign one of the lowest ranks as the reference in applying various HW transitions.

Then, several questions arise naturally.
Do the duality cascades always end regardless of the initial brane configuration?
Is the endpoint of duality cascades uniquely determined from the initial brane configuration?
Fortunately, we have found that the questions can be studied from the geometrical viewpoint of parallelotopes.
Namely, we define the fundamental domain to contain all the final destinations of duality cascades and rewrite the processes of duality cascades as translations of the fundamental domain.
Then, the above questions reduce to whether the fundamental domain can tile the whole parameter space by translations, or whether it is a parallelotope.
Besides, in the brane configurations corresponding to quantum curves with the exceptional Weyl groups, the Weyl groups help us to answer the questions.
We have found that the fundamental domain in the Weyl chamber is nothing but the affine Weyl chamber.
Then, the affine Weyl group guarantees the fundamental domain to be a parallelotope.
This result implies that the affine Weyl group may play an important role of hidden symmetries in analyzing brane configurations.

Since without deformations, only subclasses of curves are realized, which enjoy subgroups of the Weyl groups, we continue to study curves with full parameters by introducing FI parameters, though this time we do not have the HW transitions.
So we generate the whole system of inequalities by the Weyl group.
Then, we continue to find that the fundamental domain in the Weyl chamber coincides with the affine Weyl chamber.
Since we have applied the Weyl group, it is important to understand its interpretation in brane configurations.
We have identified the transformations of the Weyl group as ``halves'' of the HW transitions and presented explicit transformation rules as brane transitions \eqref{transition}.
Apparently, there are a lot of interesting future directions to pursue.

Firstly, so far we only concentrate on brane configurations with the interpretation of quantum curves enjoying symmetries of the exceptional Weyl groups.
It is interesting to discuss general brane configurations with arbitrary numbers of NS5-branes, $(1,k)$5-branes and even general $(p,q)$5-branes to understand whether they are still parallelotopes and what symmetries they have.

Secondly, in introducing FI parameters or super determinant operators and imposing the Weyl group symmetries, apparently we have introduced symmetries transforming the relative ranks with the FI parameters.
We have identified these symmetries as ``halves'' of the HW transitions and specified the expression of the transformation in \eqref{transition}.
However, the physical interpretations or the justifications from the first principle are still lacking.
We would like to understand the condition under which these symmetries hold.

Thirdly, it is interesting to point out that, in two-dimensional conformal field theories (current algebras), states and operators are also classified by representations of affine Lie algebras.
Apparently, we encounter a close similarity between our current brane configurations and two-dimensional conformal field theories.
We would like to understand whether they can be understood by bulk/boundary correspondences.

Fourthly, it is known that the time evolution in non-autonomous $q$-Painlev\'e equations is realized by the translations in the affine Weyl group.
Here, we have found that duality cascades are also realized by translations.
By combining these facts, we may want to conclude that the duality cascades are described by the Painlev\'e equations.
It is interesting to elaborate the relations between them.

Finally, there are many generalizations including the introduction of orientifold planes \cite{H2,MS1,MS2,MN5}.
It is interesting to figure out brane configurations and brane transitions with the orientifold planes.

\appendix

\section{Quantum curves and Weyl groups}

In this appendix we summarize several known results on the $D_5$ and $E_7$ quantum curves (see also \cite{KNY}).
We start with the simpler $D_5$ case and then proceed to the more complicated $E_7$ case with degeneracies next.

\subsection{$D_5$}\label{appendixD5}

\begin{figure}[!t]
\centering\includegraphics[scale=0.7,angle=-90]{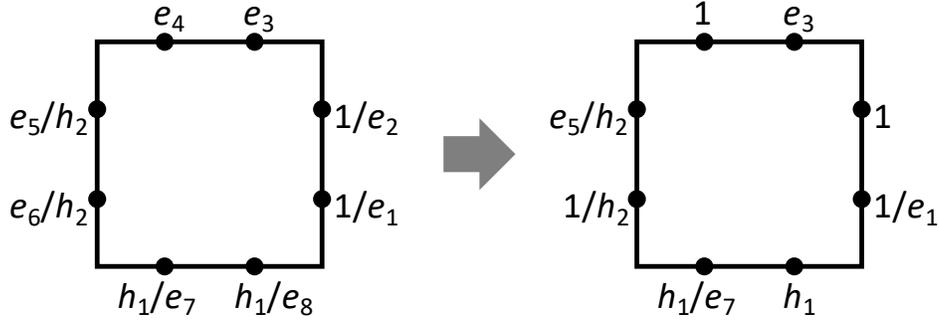}
\caption{Asymptotic values for the $D_5$ curve: the original asymptotic values (left) and the values after fixing the conditions in \eqref{d5constraint} (right).}
\label{d5config}
\end{figure}

The $D_5$ quantum curve is given by
\begin{align}
&\widehat H/\alpha=q^{-\frac{1}{2}}\widehat Q^{-1}(\widehat Q+q^{\frac{1}{2}}e_3)(\widehat Q+q^{\frac{1}{2}}e_4)\widehat P
+(e_1^{-1}+e_2^{-1})\widehat Q+E/\alpha\nonumber\\
&\quad+e_3e_4h_2^{-1}(e_5+e_6)\widehat Q^{-1}
+q^{\frac{1}{2}}(e_1e_2)^{-1}\widehat Q^{-1}(\widehat Q+q^{-\frac{1}{2}}h_1e_7^{-1})(\widehat Q+q^{-\frac{1}{2}}h_1e_8^{-1})\widehat P^{-1},
\label{d5curve}
\end{align}
with the constant $E$, the overall factor $\alpha$ and the parameters $h_1,h_2,e_1,\cdots,e_7$ subject to the constraint $(h_1h_2)^2=\prod_{i=1}^8e_i$.
Note that the factors $q$ in \eqref{d5curve} are introduced from the Weyl ordering of the quantum operators, with which the transformations are expressed succinctly.
The classical cousin of this quantum curve is characterized by the asymptotic values $(Q,P)=(\infty,-1/e_{i(=1,2)})$, $(-e_{i(=3,4)},\infty)$, $(0,-e_{i(=5,6)}/h_2)$, $(-h_1/e_{i(=7,8)},0)$.
See figure \ref{d5config}.
We symbolically denote the asymptotic values for the figure as
\begin{align}
(e_1^{-1},e_2^{-1};e_3,e_4;h_2^{-1}e_5,h_2^{-1}e_6;h_1e_7^{-1},h_1e_8^{-1}).
\label{D5asympt}
\end{align}

\begin{figure}[!t]
\centering\includegraphics[scale=0.6,angle=-90]{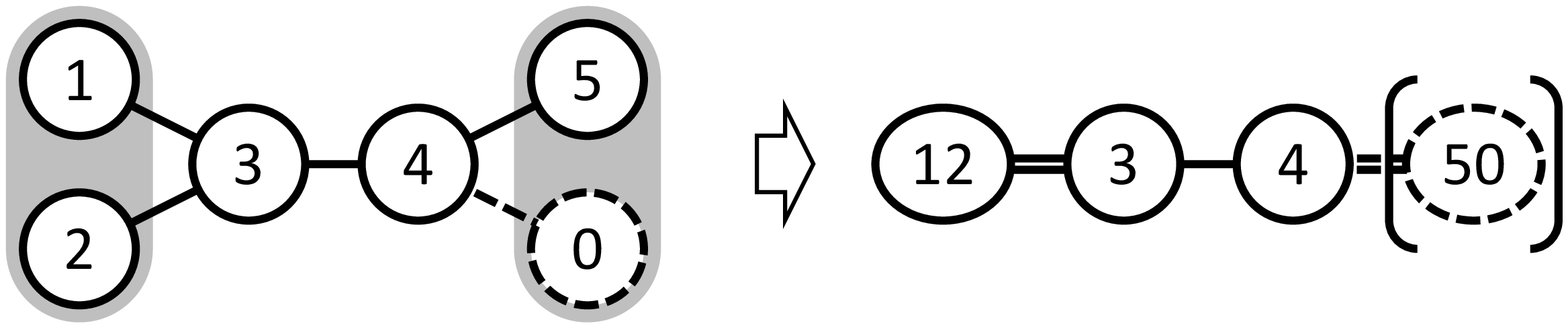}
\caption{The $\widehat D_5$ Dynkin diagram and the $B_3$ Dynkin diagram obtained by the ${\mathbb Z}_2$ folding.
In the three-dimensional subspace corresponding to brane configurations without deformations of FI parameters, only the combined transformations $s_{12}=s_1s_2$ and $s_{50}=s_5s_0$ remain as symmetries.
These transformations correspond to the trivial HW transitions for 5-branes of the same type.}
\label{d5dynkin}
\end{figure}

Apparently, the quantum curve enjoys the symmetries exchanging the asymptotic values,
\begin{align}
s_1:\frac{h_1}{e_7}\leftrightarrow\frac{h_1}{e_8},\quad
s_2:e_3\leftrightarrow e_4,\quad
s_5:\frac{1}{e_1}\leftrightarrow\frac{1}{e_2},\quad
s_0:\frac{e_5}{h_2}\leftrightarrow\frac{e_6}{h_2}.
\label{s1250he}
\end{align}
Besides, due to non-trivial canonical transformations, we have symmetries
\begin{align}
s_3:e_3\leftrightarrow\frac{h_1}{e_7},\quad
s_4:\frac{1}{e_1}\leftrightarrow\frac{e_5}{h_2}.
\label{s34he}
\end{align}
Since we are labeling eight asymptotic values with ten parameters and, by the similarity transformations, we are free to change the asymptotic values linearly in the additive notation, we can fix four parameters and solve the constraint explicitly for $e_7$ as
\begin{align}
e_2=e_4=e_6=e_8=1,\quad e_7=(h_1h_2)^2/(e_1e_3e_5).
\label{d5constraint}
\end{align}
Then, the quantum curve is parameterized by $(h_1,h_2,e_1,e_3,e_5)$.
After fixing the conditions \eqref{d5constraint}, the transformations \eqref{s1250he} and \eqref{s34he} are given by
\begin{align}
s_1&:(h_1,h_2,e_1,e_3,e_5)\mapsto\big((h_1h_2^2)^{-1}e_1e_3e_5,h_2,e_1,e_3,e_5\big),\nonumber\\
s_2&:(h_1,h_2,e_1,e_3,e_5)\mapsto\big(h_1e_3^{-1},h_2,e_1,e_3^{-1},e_5\big),\nonumber\\
s_3&:(h_1,h_2,e_1,e_3,e_5)\mapsto\big(h_1,(h_1h_2)^{-1}e_1e_5,e_1,(h_1h_2^2)^{-1}e_1e_3e_5,e_5\big),\nonumber\\
s_4&:(h_1,h_2,e_1,e_3,e_5)\mapsto\big(h_1h_2(e_1e_5)^{-1},h_2,h_2e_5^{-1},e_3,h_2e_1^{-1}\big),\nonumber\\
s_5&:(h_1,h_2,e_1,e_3,e_5)\mapsto\big(h_1,h_2e_1^{-1},e_1^{-1},e_3,e_5\big),\nonumber\\
s_0&:(h_1,h_2,e_1,e_3,e_5)\mapsto\big(h_1,h_2e_5^{-1},e_1,e_3,e_5^{-1}\big).
\label{d5transform}
\end{align}
These transformations form the $D_5$ Weyl group.
See the $\widehat D_5$ Dynkin diagram in figure \ref{d5dynkin} for the numbering of the transformations.

Then, it was found in \cite{KM} that the space of brane configurations in \eqref{M123} (with three relative ranks) is identified to be the three-dimensional subspace constrained by
\begin{align}
e_3=(h_1h_2)^2/(e_1e_3e_5),\quad
e_1=e_5.
\label{d5antipode}
\end{align}
In \cite{FMN} the relation \eqref{d5antipode} was further interpreted from the inverse relation of the asymptotic values.
Namely, in obtaining the asymptotic values, the commutation relations
\begin{align}
&\widehat P^{\frac{n}{2}}
(\widehat Q^{\frac{1}{2}}+\widehat Q^{-\frac{1}{2}})
=(q^{-\frac{n}{4}}\widehat Q^{\frac{1}{2}}+q^{\frac{n}{4}}\widehat Q^{-\frac{1}{2}})
\widehat P^{\frac{n}{2}},\nonumber\\
&(\widehat P^{\frac{1}{2}}+\widehat P^{-\frac{1}{2}})
\widehat Q^{\frac{n}{2}}
=\widehat Q^{\frac{n}{2}}
(q^{-\frac{n}{4}}\widehat P^{\frac{1}{2}}+q^{\frac{n}{4}}\widehat P^{-\frac{1}{2}}),
\label{commutation}
\end{align}
automatically guarantee that the asymptotic values for large/small $Q$ or $P$ behave inversely
\begin{align}
Ae_i\cdot A\frac{h_1}{e_{i+4}}=1\;\;\;(i=3,4),\quad
B\frac{1}{e_i}\cdot B\frac{e_{i+4}}{h_2}=1\;\;\;(i=1,2),
\label{d5anti}
\end{align}
before fixing the parameters to \eqref{d5constraint}, which reduce to the relations \eqref{d5antipode}.
We can parameterize the subspace of \eqref{d5antipode} by
\begin{align}
(h_1,h_2,e_1,e_3,e_5)
=\bigg(\frac{m_2m_3}{m_1},\frac{m_1m_3}{m_2},\frac{m_3}{m_2},m_2m_3,\frac{m_3}{m_2}\bigg),
\label{hem}
\end{align}
where the parameters are identified with the relative ranks $(M_1,M_2,M_3)$ by $m_i=e^{2\pi iM_i}$ \cite{KM}.
Then, the parameters enjoy only a subgroup of the $D_5$ Weyl group
\begin{align}
s_{12}&:(m_1,m_2,m_3)\mapsto\big(m_1,m_3^{-1},m_2^{-1}\big),\nonumber\\
s_3&:(m_1,m_2,m_3)\mapsto\bigg(\frac{1}{m_2m_3},\sqrt{\frac{m_2}{m_1m_3}},\sqrt{\frac{m_3}{m_1m_2}}\bigg),\nonumber\\
s_4&:(m_1,m_2,m_3)\mapsto\bigg(\frac{m_3}{m_2},\sqrt{\frac{m_2m_3}{m_1}},\sqrt{m_1m_2m_3}\bigg),\nonumber\\
s_{50}&:(m_1,m_2,m_3)\mapsto\big(m_1,m_3,m_2\big),
\label{B3Weyl}
\end{align}
which forms the $B_3$ Weyl group.
Namely, in the subspace of the brane configurations without FI parameters, trivial exchanges of the asymptotic values $s_1$, $s_2$, $s_5$, $s_0$ have to combine into $s_{12}=s_1s_2$ and $s_{50}=s_5s_0$, which are interpreted as the trivial HW transitions exchanging 5-branes of the same type.
These generate the $B_3$ Weyl group after including the transformations $s_3$ and $s_4$.

Indeed, the transformations \eqref{B3Weyl} can be interpreted as Weyl reflections associated to the simple roots $\alpha$,
\begin{align}
s_\alpha\lambda=\lambda-2\alpha\frac{(\alpha,\lambda)}{(\alpha,\alpha)},
\label{reflection}
\end{align}
by changing the multiplicative variables $m_i$ into the additive ones $M_i$ with $m_i=e^{2\pi iM_i}$,
\begin{align}
s_{12}&:(M_1,M_2,M_3)\mapsto(M_1,-M_3,-M_2),\nonumber\\
s_3&:(M_1,M_2,M_3)\mapsto(-M_2-M_3,(-M_1+M_2-M_3)/2,(-M_1-M_2+M_3)/2),\nonumber\\
s_4&:(M_1,M_2,M_3)\mapsto(-M_2+M_3,(-M_1+M_2+M_3)/2,(M_1+M_2+M_3)/2),\nonumber\\
s_{50}&:(M_1,M_2,M_3)\mapsto(M_1,M_3,M_2),
\label{sM}
\end{align}
though the reflection does not necessarily respect the orthonormal Euclidean metric.
For the interpretation, we need to choose the metric for the bilinear form $(*,*)$ in the parameter space $\{\lambda=(M_1,M_2,M_3)\}$ and the simple roots to be $g=\diag(1,2,2)$ and
\begin{align}
\alpha_{12}=(0,1/2,1/2),\quad
\alpha_3=(-1,-1/2,-1/2),\quad
\alpha_4=(1,1/2,-1/2),\quad
\alpha_{50}=(0,-1/2,1/2),
\label{B3roots}
\end{align}
where the dual fundamental weights $(\alpha_i,\omega_j)=\delta_{ij}$ are given by
\begin{align}
\omega_{12}=(-1/2,1/2,0),\quad
\omega_3=(-1,1/2,-1/2),\quad
\omega_4=(0,1/2,-1/2).
\label{B3fw}
\end{align}

To relate the parameters $(h_1,h_2,e_1,e_3,e_5)$ to those of relative ranks and FI parameters, we have also introduced the parameters $(m_1,m_2,m_3,z_1,z_3)$ in \eqref{hemz}.
Then, the transformations of the $D_5$ Weyl group \eqref{d5transform} in these parameters are given as
\begin{align}
s_1&:(m_1,m_2,m_3,z_1,z_3)\mapsto\bigg(m_1,\sqrt{\frac{m_2z_3}{m_3}},\sqrt{\frac{m_3z_3}{m_2}},z_1,m_2m_3\bigg),\nonumber\\
s_2&:(m_1,m_2,m_3,z_1,z_3)\mapsto\bigg(m_1,\sqrt{\frac{m_2}{m_3z_3}},\sqrt{\frac{m_3}{m_2z_3}},z_1,\frac{1}{m_2m_3}\bigg),\nonumber\\
s_3&:(m_1,m_2,m_3,z_1,z_3)\mapsto\bigg(\frac{1}{m_2m_3},\sqrt{\frac{m_2}{m_1m_3}},\sqrt{\frac{m_3}{m_1m_2}},z_1,z_3\bigg),\nonumber\\
s_4&:(m_1,m_2,m_3,z_1,z_3)\mapsto\bigg(\frac{m_3}{m_2},\sqrt{\frac{m_2m_3}{m_1}},\sqrt{m_1m_2m_3},z_1,z_3\bigg),\nonumber\\
s_5&:(m_1,m_2,m_3,z_1,z_3)\mapsto\bigg(m_1,\sqrt{\frac{m_2m_3}{z_1}},\sqrt{m_2m_3z_1},\frac{m_3}{m_2},z_3\bigg),\nonumber\\
s_0&:(m_1,m_2,m_3,z_1,z_3)\mapsto\bigg(m_1,\sqrt{m_2m_3z_1},\sqrt{\frac{m_2m_3}{z_1}},\frac{m_2}{m_3},z_3\bigg).
\label{D5transf}
\end{align}
These are also interpreted as Weyl reflections \eqref{reflection} in $\{\lambda=(M_1,M_2,M_3,Z_1,Z_3)\}$ if we choose the metric for the bilinear form and the simple roots to be $g=\diag(1,2,2,1,1)$ and
\begin{align}
&\alpha_1=(0,1/2,1/2,0,-1),\quad
\alpha_2=(0,1/2,1/2,0,1),\quad
\alpha_3=(-1,-1/2,-1/2,0,0),\nonumber\\
&\alpha_4=(1,1/2,-1/2,0,0),\quad
\alpha_5=(0,-1/2,1/2,-1,0).
\label{pointD5}
\end{align}
Here the dual fundamental weights are given by
\begin{align}
&\omega_1=(-1,1,0,-1,-1),\quad\omega_2=(-1,1,0,-1,1),\quad\omega_3=(-1/2,1/4,-1/4,-1/2,0),\nonumber\\
&\omega_4=(0,1/4,-1/4,-1/2,0),\quad\omega_5=(0,0,0,-2,0).
\label{D5fw}
\end{align}

\subsection{$E_7$}\label{appendixE7}

\begin{figure}[!t]
\centering\includegraphics[scale=0.7,angle=-90]{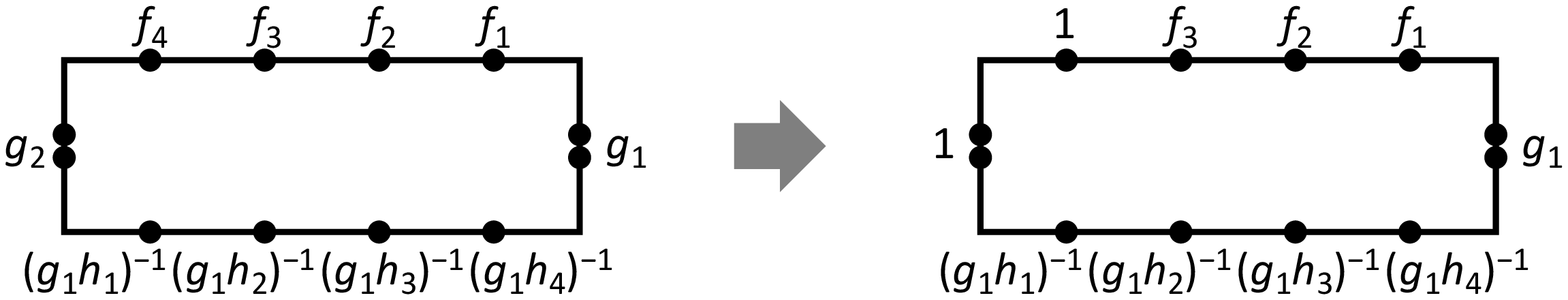}
\caption{Asymptotic values for the $E_7$ curve: the original asymptotic values (left) and the values after fixing the conditions in \eqref{fghcondition} (right).}
\label{e7config}
\end{figure}

Now let us turn to the $E_7$ case.
As noticed in \cite{FMN} the main difficulty is the degeneracy of the $E_7$ quantum curve.
The $E_7$ quantum curve is given by
\begin{align}
&\widehat H/\alpha=q^{-1}\widehat Q^{-2}\prod_{i=1}^4(\widehat Q+q^{\frac{1}{2}}f_i)\widehat P
+[2]_qg_1\widehat Q^2+({\tilde F}_1g_1+{\tilde F}_4{\tilde G}_2^2\,{\tilde H}_3)\widehat Q+E/\alpha\nonumber\\
&\qquad+({\tilde F}_3g_2+{\tilde F}_4{\tilde G}_2{\tilde H}_1)\widehat Q^{-1}+[2]_q{\tilde F}_4g_2\widehat Q^{-2}
+qg_1^2\widehat Q^{-2}\prod_{i=1}^4(\widehat Q+q^{-\frac{1}{2}}(g_1h_i)^{-1})\widehat P^{-1},
\end{align}
with the $q$-integer $[2]_q=q^{\frac{1}{2}}+q^{-\frac{1}{2}}$ and ten parameters $(f_1,f_2,f_3,f_4,g_1,g_2,h_1,h_2,h_3,h_4)$ forming 
\begin{align}
\prod_{i=1}^4(1+zf_i)=\sum_{n=0}^4{\tilde F}_nz^n,\quad
\prod_{i=1}^2(1+zg_i)=\sum_{n=0}^2{\tilde G}_nz^n,\quad
\prod_{i=1}^4(1+zh_i)=\sum_{n=0}^4{\tilde H}_nz^n,
\end{align}
subject to the constraint ${\tilde F}_4{\tilde G}_2^2{\tilde H}_4=1$.
The asymptotic values are symbolically given by 
\begin{align}
(g_1\text{[double]};f_1,f_2,f_3,f_4;g_2\text{[double]};(g_1h_1)^{-1},(g_1h_2)^{-1},(g_1h_3)^{-1},(g_1h_4)^{-1}),
\label{E7asympt}
\end{align}
(see figure \ref{e7config}).

\begin{figure}[!t]
\centering\includegraphics[scale=0.6,angle=-90]{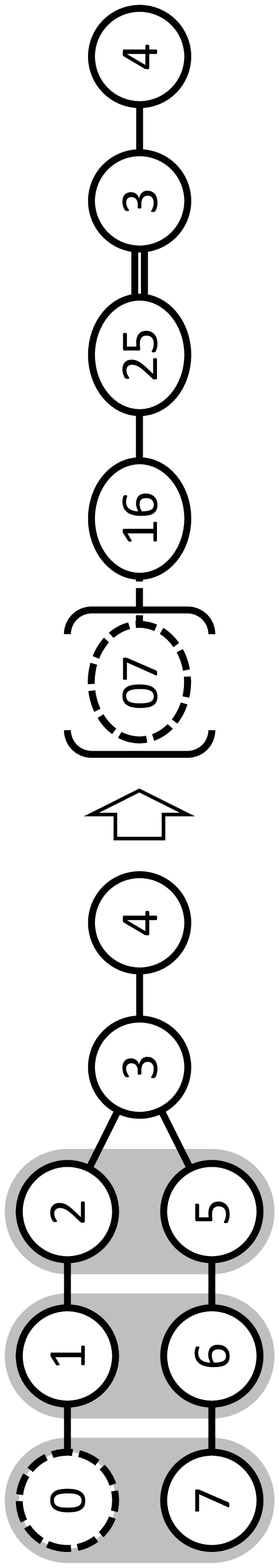}
\caption{The $\widehat E_7$ Dynkin diagram and the $F_4$ Dynkin diagram obtained by the ${\mathbb Z}_2$ folding.
In the four-dimensional subspace corresponding to brane configurations without deformations of FI parameters, only the combined transformations $s_{07}=s_0s_7$, $s_{16}=s_1s_6$ and $s_{25}=s_2s_5$ remain as symmetries.
These transformations correspond to the trivial HW transitions for 5-branes of the same type.}
\label{e7dynkin}
\end{figure}

By fixing extra degrees of freedom and solving for the constraint
\begin{align}
f_4=g_2=1,\quad
h_4=(f_1f_2f_3g_1^2h_1h_2h_3)^{-1},
\label{fghcondition}
\end{align}
we end up with seven parameters $(f_1,f_2,f_3,g_1,h_1,h_2,h_3)$.
Then, it was known that the curve enjoys symmetries of the $E_7$ Weyl group generated by
\begin{align}
s_0&:(f_1,f_2,f_3,g_1,h_1,h_2,h_3)
\mapsto\big(f_3^{-1}f_1,f_3^{-1}f_2,f_3^{-1},g_1,f_3h_1,f_3h_2,f_3h_3\big),\nonumber\\
s_1&:(f_1,f_2,f_3,g_1,h_1,h_2,h_3)
\mapsto\big(f_1,f_3,f_2,g_1,h_1,h_2,h_3\big),\nonumber\\
s_2&:(f_1,f_2,f_3,g_1,h_1,h_2,h_3)
\mapsto\big(f_2,f_1,f_3,g_1,h_1,h_2,h_3\big),\nonumber\\
s_3&:(f_1,f_2,f_3,g_1,h_1,h_2,h_3)
\mapsto\big((g_1h_1)^{-1},f_2,f_3,(f_1h_1)^{-1},h_1,(f_1g_1h_1)h_2,(f_1g_1h_1)h_3\big),\nonumber\\
s_4&:(f_1,f_2,f_3,g_1,h_1,h_2,h_3)
\mapsto\big(f_1,f_2,f_3,g_1^{-1},g_1h_1,g_1h_2,g_1h_3\big),\nonumber\\
s_5&:(f_1,f_2,f_3,g_1,h_1,h_2,h_3)
\mapsto\big(f_1,f_2,f_3,g_1,h_2,h_1,h_3\big),\nonumber\\
s_6&:(f_1,f_2,f_3,g_1,h_1,h_2,h_3)
\mapsto\big(f_1,f_2,f_3,g_1,h_1,h_3,h_2\big),\nonumber\\
s_7&:(f_1,f_2,f_3,g_1,h_1,h_2,h_3)
\mapsto\big(f_1,f_2,f_3,g_1,h_1,h_2,(f_1f_2f_3g_1^2h_1h_2h_3)^{-1}\big).
\label{e7transformation}
\end{align}
See figure \ref{e7dynkin} for the $\widehat E_7$ Dynkin diagram and the numbering of the transformations.

For the interpretation in the brane configurations \eqref{F4braneconfig} without deformations, we have to impose the relation
\begin{align}
Af_i\cdot A(g_1h_i)^{-1}=1\;\;\;(i=1,2,3,4),\quad Bg_2\cdot Bg_1=1.
\end{align}
As in the $D_5$ case, due to the relation of the asymptotic values in the opposite regimes, only the combined transformations $s_{07}=s_0s_7$, $s_{16}=s_1s_6$, $s_{25}=s_2s_5$ remain as symmetries.
Namely, in terms of the parameters
\begin{align}
(f_1,f_2,f_3,g_1,h_1,h_2,h_3)
=\bigg(f_1,f_2,f_3,g_1,\sqrt{\frac{f_1}{f_2f_3g_1}},\sqrt{\frac{f_2}{f_1f_3g_1}},\sqrt{\frac{f_3}{f_1f_2g_1}}\bigg),
\label{F4subspace}
\end{align}
the transformations are given by
\begin{align}
s_{07}&:(f_1,f_2,f_3,g_1)\mapsto\big(f_1f_3^{-1},f_2f_3^{-1},f_3^{-1},g_1\big),\nonumber\\
s_{16}&:(f_1,f_2,f_3,g_1)\mapsto\big(f_1,f_3,f_2,g_1\big),\nonumber\\
s_{25}&:(f_1,f_2,f_3,g_1)\mapsto\big(f_2,f_1,f_3,g_1\big),\nonumber\\
s_3&:(f_1,f_2,f_3,g_1)\mapsto
\bigg(\sqrt{\frac{f_2f_3}{f_1g_1}},f_2,f_3,\sqrt{\frac{f_2f_3g_1}{f_1^3}}\bigg),\nonumber\\
s_4&:(f_1,f_2,f_3,g_1)\mapsto\big(f_1,f_2,f_3,g_1^{-1}\big).
\label{F4Weyl}
\end{align}
As in \eqref{B3Weyl}, the combined transformations are interpreted as the trivial HW transitions for 5-branes of the same type.

To interpret the transformations as Weyl reflections, we need to rewrite the transformations \eqref{F4Weyl} for the multiplicative variables into those for the additive ones by $(f_1,f_2,f_3,g_1)=(e^{2\pi iF_1},e^{2\pi iF_2},e^{2\pi iF_3},e^{2\pi iG_1})$.
Then, we can interpret these transformations as Weyl reflections in $\{\lambda=(F_1,F_2,F_3,G_1)\}$ by choosing the metric for the bilinear form
\begin{align}
g=\begin{pmatrix}\frac{3}{2}&-\frac{1}{2}&-\frac{1}{2}&0\\
-\frac{1}{2}&\frac{3}{2}&-\frac{1}{2}&0\\
-\frac{1}{2}&-\frac{1}{2}&\frac{3}{2}&0\\
0&0&0&\frac{1}{2}\end{pmatrix},
\label{F4metric}
\end{align}
and the simple roots
\begin{align}
&\alpha_{07}=(1/2,1/2,1,0),\quad
\alpha_{16}=(0,1/2,-1/2,0),\quad
\alpha_{25}=(1/2,-1/2,0,0),\nonumber\\
&\alpha_3=(-1,0,0,-1),\quad
\alpha_4=(0,0,0,2).
\label{F4sr}
\end{align}
Here the dual fundamental weights are given by
\begin{align}
&\omega_{16}=(-1/2,-1/2,-1,0),\quad
\omega_{25}=(-1,-3/2,-3/2,0),\nonumber\\
&\omega_3=(-2,-2,-2,0),\quad
\omega_4=(-1,-1,-1,1).
\label{F4fw}
\end{align}

Note that we can also interpret the symmetries \eqref{e7transformation} as Weyl reflections \eqref{reflection} in $\{\lambda=(F_1,F_2,F_3,G_1,H_1,H_2,H_3)\}$ by choosing the metric for the bilinear form and the simple roots to be
\begin{align}
g=\begin{pmatrix}
\frac{3}{2}&\frac{1}{2}&\frac{1}{2}&\frac{3}{2}&1&1&1\\
\frac{1}{2}&\frac{3}{2}&\frac{1}{2}&\frac{3}{2}&1&1&1\\
\frac{1}{2}&\frac{1}{2}&\frac{3}{2}&\frac{3}{2}&1&1&1\\
\frac{3}{2}&\frac{3}{2}&\frac{3}{2}&\frac{7}{2}&2&2&2\\
1&1&1&2&2&1&1\\
1&1&1&2&1&2&1\\
1&1&1&2&1&1&2
\end{pmatrix},
\end{align}
and
\begin{align}
&\alpha_1=(0,1,-1,0,0,0,0),\quad\alpha_2=(1,-1,0,0,0,0,0),\nonumber\\
&\alpha_3=(-1,0,0,-1,0,1,1),\quad\alpha_4=(0,0,0,2,-1,-1,-1),\nonumber\\
&\alpha_5=(0,0,0,0,1,-1,0),\quad\alpha_6=(0,0,0,0,0,1,-1),\quad\alpha_7=(0,0,0,0,0,0,1).
\end{align}
Here the dual fundamental weights are given by
\begin{align}
&\omega_1=(-1,-1,-2,0,1,1,1),\quad
\omega_2=(-2,-3,-3,0,2,2,2),\nonumber\\
&\omega_3=(-4,-4,-4,0,3,3,3),\quad
\omega_4=(-2,-2,-2,1,1,1,1),\nonumber\\
&\omega_5=(-3,-3,-3,0,3,2,2),\quad
\omega_6=(-2,-2,-2,0,2,2,1),\quad
\omega_7=(-1,-1,-1,0,1,1,1).
\label{E7fw}
\end{align}

\section*{Acknowledgements}
We are grateful to Naotaka Kubo, Tomoki Nosaka, Hikaru Sasaki and Yasuhiko Yamada for valuable discussions and comments.
The work of T.F. and S.M. is supported respectively by Grant-in-Aid for JSPS Fellows \#20J15045 and Grant-in-Aid for Scientific Research (C) \#19K03829.
S.M.\ would like to thank Yukawa Institute for Theoretical Physics at Kyoto University for warm hospitality.

\end{document}